\documentclass[aps,floatfix,epsfig]{revtex4}
\usepackage{pdfpages}
\usepackage{subfigure}
\usepackage{amsmath}
\usepackage{eqnarray,amsmath}
\usepackage{amssymb}
\usepackage{amsthm}
\usepackage{epstopdf}
\usepackage{mathrsfs}
\usepackage{natbib}
\usepackage{amssymb,latexsym}
\usepackage{dcolumn}
\usepackage{graphicx}
\usepackage{multirow}
\def\be{\begin{equation}}
	\def\ee{\end{equation}}
\def\ba{\begin{eqnarray}}
	\def\ea{\end{eqnarray}}

\begin{document}
\title{The Effect of Interacting Dark Energy on Mass-Temperature Relation in Galaxy Clusters}

\author{Mahdi Naseri}
\email{mahdi.naseri@email.kntu.ac.ir}
\affiliation{Department of Physics, K.N. Toosi University of Technology, P. O. Box 15875-4416, Tehran, Iran}

\author{Javad T. Firouzjaee}
\email{firouzjaee@kntu.ac.ir}
\affiliation{Department of Physics, K.N. Toosi University of Technology, P. O. Box 15875-4416, Tehran, Iran }
\affiliation{ School of physics, Institute for Research in Fundamental Sciences (IPM), P. O. Box 19395-5531, Tehran, Iran }

\begin{abstract}

\textbf{Abstract:} There are a variety of cosmological models for dark matter and dark energy in which a possible interaction is considered between these two significant components of the universe. We focus on five suggested models of interacting dark matter and dark energy and derive the modified virial theorem for them by developing a previous approach. It provides an opportunity to study the evolution of this modified virial theorem with time and interacting constants for different interacting models. Then we use this obtained virial condition to investigate the modified mass-temperature relation in galaxy clusters via three various methods. It reveals that the effect of interaction between dark matter and dark energy merely appears in the normalization factor of $M\propto T^{\frac{3}{2}}$. This relation also leads to a new constraint on the constants of interacting models, which only depends on the concentration parameter and density profile of the cluster. Then we use five observational data sets to check some proposed figures for the constants of interaction which have been resulted from other observational constraints. Finally, by fitting the observational results to the modified mass-temperature relation, we obtain values for interacting constants of three models and four specific cases of the two remained models. In agreement with many other observational outcomes, we find that according to observational data for masses and temperatures of the galaxy clusters, energy transfer occurs from dark matter to dark energy in the seven investigated models. 

\end{abstract}
%
%

\maketitle

\tableofcontents
\section{Introduction}

As different observational outcomes have revealed the existence of two unfamiliar contributors to the physics of the universe, research into the "dark sector" has gained currency in modern cosmology. Dark matter (DM) proposed to clarify rotation curves of spiral galaxies, and the idea behind dark energy (DE) was initially formed to explain the late-time acceleration of the universe. Eventually, the $\Lambda$CDM model accounted for the primary suggestion for the cosmos. \\

In spite of gravitational evidence for DM from galaxies \cite{DM-galaxy}, clusters of galaxies \cite{cluster-ga}, cosmic microwave background (CMB) anisotropies \cite{anisotropies}, cosmic shear \cite{shear}, structure formation \cite{structure-formation} and large-scale structure of the Universe \cite{large-structure}, last years of direct and indirect searches of those DM particles did not give any convinced result \cite{large-scale}. In addition, the accelerated expansion of the universe modeled with $\Lambda$ \cite{lamda} raised several problems, including the "cosmological constant finetuning problem" and the "cosmic coincidence problem" \cite{cosmic-coincidence}. \\

 However, it could be possible to assume and investigate more elaborate alternatives in which there is a feasible non-gravitational interaction between DM and DE. The idea has extended in \cite{Farrar:2003uw}, where DM particle mass is determined according to its interaction with a scalar field with the energy density of DE. Such an assumption resembles how the Higgs field results in quark and lepton masses via interacting with them.\\
 
Not only is the notion of interacting dark sector interesting, but it could also be beneficial in terms of solving some cosmological problems. By way of illustration, it may explain why the densities of DE and DM are of the same order, despite the fact that they evolve differently with redshift, namely the "coincidence problem" (see e.g. \cite{CalderaCabral:2008bx}). The interacting dark energy model should justify the same observation in contrast to the $\Lambda$CDM model which modified gravity models do \cite{interacting-observation,vonMarttens:2018iav}. \\

One can study the effects of modified gravity with structure formation and verified employing dark-matter-only N-body simulations
\cite{N-body-simulations}.  Since experiments only measure
photons which are emitted from the baryonic matter, photons properties cannot be directly calculated only from dark matter simulations. However, hydrodynamical simulations are more appropriate in the observational aspect, as they provide
observables, such as the halo profile, the turnaround radius \cite{turnaround},
the splashback radius \cite{splashback} and the mass-temperature (M-T) relation
 \cite{Hammami-16-74}. \\ 

There are a wide range of observations, simulations, and theoretical research into the relationship between mass and temperature of galaxy clusters which have been done heretofore. The only consensus among all these endeavors is admit of an evident correlation between the total gravitational mass of the clusters, X-Ray luminosity, and thereby, their temperature (that is the temperature of the intracluster medium i.e ICM). It is of significance to study this relation, owing to the fact that the cluster masses are arduous to measure directly through observation. Fundamental arguments based on virialization density suggest that $M\propto T^{\frac{3}{2}}$, where T is the temperature of a cluster within a
certain radius (e.g. the virial radius) and M is the mass
within the same radius (see \cite{Afshordi:2001ze, Popolo} for advanced discussion). The mass-temperature relation can be directly compared with observations. This relation has been used to put constraints on modified gravity models. For example, using the hydrodynamical simulations,  \cite{Hammami-16-74} showed that the M-T relation obtained
in modified gravity theories is different from the expectations of the general relativity. Nevertheless, \cite{DelPopolo:2019oxn} showed that the mass-temperature relation of the $\Lambda$CDM model is similar to that of the f(R) and symmetron models.\\

The paper is organized as follows. Section II briefly presents the interacting dark energy model and specifically, introduces five interacting models on which we concentrate in this study. We also obtain the virial theorem for these interacting models. Section III is devoted to the mass-temperature relation of galaxy clusters concerning the interaction between dark matter and dark energy. Section IV makes a comparison between observational data and obtained M-T relation to study constants of interaction in the five models. We summarize and give our final thoughts in Section V.\\

\section{Interacting Dark Energy Models and Virial Theorem}

The interacting dark energy model is composed of dark matter and dark energy only, as a flat Friedmann-Lemaitre-Robertson-Walker (FLRW) background metric. The dark sector interaction is modeled with a heat flux in the Bianchi identities between the two dark components as
\be
\nabla_\mu T_{(\lambda)}^{\mu\nu} \ne 0 \, ,
\ee
where $T_{(\lambda)}^{\mu\nu}$ in the energy-momentum tensor of each individual component which is no longer conserved. \\

There are a number of interacting models which have been suggested and investigated recently. According to \cite{CalderaCabral:2008bx}, the balance, Raychaudhuri and FLRW equations can be written as
\begin{equation} \label{eq:motiond}
\dot{\rho}_b = - 3H \rho_b \, ,
\end{equation}
\begin{equation} \label{eq:motiona}
\dot{\rho}_c = - 3H \rho_c + Q \, ,
\end{equation}
\begin{equation} \label{eq:motionb}
  \dot{\rho}_x = - 3(1+w_x) H \rho_x - Q \, ,
\end{equation}
\begin{equation} \label{eq:motionc}
  \dot{H} = - 4\pi G \left[ \rho_b + \rho_c + (1+w_x) \rho_x 
    \right] \, ,
\end{equation}
\begin{equation} \label{eq:friedmann}
H^2 = \frac{8\pi G}{3} \left( \rho_b + \rho_c + \rho_x \right) \, ,
\end{equation}
where $H$ is the Hubble parameter, $\rho_c$ is the cold dark matter density, $\rho_b$ is the baryonic matter density and $\rho_x$ represents the density of dark energy (with $w_x <0$ constant of its equation of state (EOS)).\\

Here, $Q$ describes the rate of energy density transfer between DE and DM, which is resulted from the interaction between them. For $Q > 0$, it describes the transfer of energy from DE to DM, and on the other hand, $Q < 0$ shows the transfer of energy from DM to DE. Note that baryons ($b$) and photons ($\gamma$), are not coupled to the dark sector; therefore, $Q_\gamma$ and $Q_b$ considered to be equal to zero.\\

A variety of functions have been proposed and studied for $Q$, including linear and non-linear combinations of  $\rho_x$ and $\rho_c$. In this paper, we concentrate on five various models for $Q$, which are rather simple and common in literature:
\begin{eqnarray}\label{Models}
\mbox{ Model I} : Q = 3H (\alpha_c \rho_c + \alpha_x \rho_x) \, ,  \nonumber \\ \nonumber \\
\mbox { Model II} : Q = 3H \xi_1 \frac{\rho_c\rho_x}{\rho_c+\rho_x} \, ,  \nonumber \\ \nonumber \\
\mbox{ Model III} : Q = 3H \xi_2 \frac{\rho_x^2}{\rho_c+\rho_x} \, ,  \nonumber \\ \nonumber \\
\mbox{ Model IV} : Q = 3H \xi_3 \frac{\rho_c^2}{\rho_c+\rho_x} \, ,  \nonumber \\ \nonumber \\
\mbox{ Model V} : Q = 3 (\Gamma_c \rho_c + \Gamma_x \rho_x) \, .
\end{eqnarray}
Here, $\xi_1$, $\xi_2$, $\xi_3$, $\alpha_j$ and $\Gamma_j$ are the main parameters of interacting dark sector ($j=c,x$). First four models are interesting, due to being coefficient with the Hubble parameter, which leads to more straightforward calculations. Whereas, Model V is more complicated and has a physical meaning. According to this model, the oscillation inflaton field decays into relativistic particles during reheating process after inflation in early universe, and $\Gamma_j$ describes decay width \cite{CalderaCabral:2008bx}. Constant parameters in Models I to IV are dimensionless, while in Model V, $\Gamma_j$ has the dimension of the Hubble parameter. For further explanations about these choices for $Q$, look at \cite{CalderaCabral:2009ja} and \cite{vonMarttens:2018iav}. \\

\subsection{Virial Theorem in Interacting Models}
In any theory of modified gravity, the virial theorem may significantly change from its Newtonian form.
To find a virial relation in the context of general relativity, one has to use the covariant collisionless Boltzmann
equation (see \cite{grvirial} and reference therein).   This approach has been extended to the virial theorem in the modified gravity theories to study the dynamics of clusters of galaxies \cite{grvirialMG}. In homogeneous and isotropic background in which gravity is not strong, the virial theorem gets the Newtonian form. \\

Before analyzing mass-temperature relation in galaxy clusters, we have to investigate modifications to the virial theorem with regard to interacting dark sector. In order to achieve this objective, we derive the Layser-Irvine equation for Models I to V and then use this equation to obtain the virial condition. This equation, and hence the virial theorem, has been driven in \cite{He:2010ta} for Model I; however, we re-write calculations so as to check it for the other four models, as well. \\

Considering Model V, the perturbation equations for DE and DM in the subhorizon scale, which have been driven in \cite{He:2009mz}, can be written in the real space as
\begin{equation} \label{DM1}
\Delta_c'+\nabla_{\bar{r}}\cdot v_c= 3\Gamma_x(\Delta_x-\Delta_c)/R \, ,
\end{equation}
\begin{equation} \label{DM2}
v_c'+\mathcal{H}v_c=-\nabla_{\bar{r}}\Psi-3(\Gamma_c+\Gamma_x/R)v_c \, .
\end{equation}
Here, $\mathcal{H}$ indicates the Hubble parameter in the conformal time, $v_c$ represents velocity of dark matter element, $\bar{r}$ refers to conformal coordinates and the prime denotes the derivative with respect to conformal time. Density contrasts of DM and DE are defined as $\Delta_c\approx \delta \rho_c/\rho_c=\delta_c$ and $\Delta_x\approx \delta \rho_x/\rho_x=\delta_x$, and we symbolize dark energy to dark matter (DM-DE) ratio by $R=\rho_c/\rho_x$. Moreover, $\Psi=\psi_m+\psi_d$ is the peculiar potential and is described by Poisson equation:
\begin{equation} \label{poisson}
\nabla^2\psi_j=4\pi G (1+3w_j) \delta \rho_j \, ,
\end{equation}
where "$j$" stands for DM or DE. Considering $\nabla_{r}=\frac{1}{a}\nabla_{\bar{r}}$ and defining $\sigma_c=\delta \rho_c$ and $\sigma_x=\delta \rho_x$, Eqs. (\ref{DM1}) and (\ref{DM2}) can be written as
\begin{equation} \label{DM11}
\dot{\sigma}_c+3H\sigma_c+\nabla_{r}(\rho_cv_c)=3(\Gamma_c\sigma_c+\Gamma_x\sigma_x) \, ,
\end{equation}
\begin{equation} \label{DM22}
\frac{\partial}{\partial t}(a v_c)=-\nabla_r (a\psi_c + a\psi_x)-3(\Gamma_c+\Gamma_x/R)(av_c) \, ,
\end{equation}
where $a$ is background scale factor and $H$ is its Hubble parameter. Following the method of \cite{He:2010ta} and \cite{Layzer}, we multiply both sides of Eq. (\ref{DM22}) by $av_c\rho_c\hat{\varepsilon}$ and then integrate them over the volume ("$\hat{\varepsilon}$" indicates volume element with criterion of expansion $\frac{\partial}{\partial t}\hat{\varepsilon}=3H\hat{\varepsilon}$ ). For the left-hand side of Eq. (\ref{DM22}), it is possible to write:
\begin{equation} \label{LHS} 
\int av_c\frac{\partial}{\partial t}(av_c)\rho_c\hat{\varepsilon} = \int av_c(\dot{a}v_c + a\dot{v_c})\rho_c\hat{\varepsilon} = \int a^2 H \rho_c v_c^2 \hat{\varepsilon} + \int a^2 \rho_c v_c \dot{v_c} \hat{\varepsilon} \, .
\end{equation}
The kinetic energy "$K_c$", which stems from the movement of DM particles, is defined as:
\begin{equation} \label{kinetik} 
K_c=\frac{1}{2}\int v_c^2 \rho_c \hat{\varepsilon} \, .
\end{equation}
It is possible to use this definition and write:
\begin{equation} \label{C} 
\frac{\partial}{\partial t}\left(a^2K_c\right)=2a\dot{a}K_c + a^2 \frac{\partial}{\partial t}K_c 
= 2a^2 HK_c + a^2 [\int v_c \dot{v_c} \rho_c \hat{\varepsilon} + \frac{1}{2}\int v_c^2 \dot{\rho_c} \hat{\varepsilon}  + \frac{1}{2} 3H \int v_c^2 \rho_c \hat{\varepsilon}] \, .
\end{equation}
Using Eq. (\ref{C}) in Eq. (\ref{LHS}) we have:
\begin{equation} \label{LHS2} 
\int av_c\frac{\partial}{\partial t}(av_c)\rho_c\hat{\varepsilon} = \frac{\partial}{\partial t}\left(a^2K_c\right) - \frac{1}{2} a^2 \int v_c^2 \dot{\rho_c} \hat{\varepsilon} - \frac{1}{2}3H a^2 \int v_c^2 \rho_c \hat{\varepsilon} \, .
\end{equation}
Then, using Eq.~(\ref{eq:motiona}) with $Q$ of the Model V in the last equation gives:
\begin{equation} \label{LHSf} 
\int av_c\frac{\partial}{\partial t}(av_c)\rho_c\hat{\varepsilon} = \frac{\partial}{\partial t}\left(a^2K_c\right) - 3a^2 (\Gamma_c + \Gamma_x/R) K_c \, .
\end{equation}
For the first term in the right-hand side of Eq.~(\ref{DM22}), integration gives:
\ba \label{RHS1} 
-\int av_c \nabla_{r}(a \psi_c+a\psi_x) \rho_c \hat{\varepsilon}=  a^2 \int \nabla_{r}(\rho_c v_c)\psi_c \hat{\varepsilon}+a^2 \int \nabla_{r}(\rho_c v_c)\psi_x \hat{\varepsilon}\, .
\ea
With the aid of Eq.~(\ref{DM11}), it can be related to potential energy
\ba \label{RHS1f} 
& -\int av_c \nabla_{r}(a \psi_c+a\psi_x) \rho_c \hat{\varepsilon}= \nonumber \\ \nonumber \\ 
& -a^2(\dot{U}_{cc}+HU_{cc}) -a^2\int \psi_x \frac{\partial}{\partial t} (\sigma_c \hat{\varepsilon}) + 3a^2 \left\{\Gamma_cU_{cx}+\Gamma_xU_{xc}+2\Gamma_cU_{cc}+2\Gamma_xU_{xx}\right\}\, ,
\ea
where $U_{\alpha\beta}=\frac{1}{2}\int \sigma_\alpha \psi_\beta \hat{\varepsilon}$; "$\alpha$" and "$\beta$" stand for DM and DE, interchangeably. \\

Eventually, integrating the second term in the right-hand side of Eq.~(\ref{DM22}) leads to
\begin{equation} \label{RHS2f} 
-\int(av_c)^2 3(\Gamma_c+\Gamma_x/R)\rho_c\hat{\varepsilon}=-6 a^2(\Gamma_c+\Gamma_x/R)K_c \, .
\end{equation}
Now, the Layzer-Irvine equation could be easily produced by combination of Eqs.~(\ref{LHSf}), (\ref{RHS1f}) and (\ref{RHS2f}) as
\ba \label{Layzer-Irvine} 
& \dot{K}_c+\dot{U}_{cc} + H(2K_c+U_{cc})= \nonumber \\ \nonumber \\
& -\int \psi_x\frac{\partial}{\partial t}(\sigma_c \hat{\varepsilon})-3(\Gamma_c+\Gamma_x/R)K_c +3\left\{\Gamma_cU_{cx}+\Gamma_xU_{xc}+2\Gamma_cU_{cc}+2\Gamma_2U_{xx}\right\} \, .
\ea
In virial equilibrium, the first and second terms of the previous equation are equal to zero. With the assumption of homogeneous distribution of DE, $\sigma_x=0$, we get
\begin{equation} \label{virial} 
K_c=- \frac{H-6\Gamma_c}{2H+3\Gamma_c+3\Gamma_x/R} U_{cc} \, .
\end{equation}
In order to facilitate following calculations, we define parameter "$\lambda_i$" and represent the virial condition as
\begin{equation} \label{virial2}
K_c=-\lambda_i U_{cc} \, .
\end{equation}
Obviously, "$\lambda_i$" is not necessarily equal to $\frac{1}{2}$ in interacting models and depends on interaction constants within $Q$. The same procedure could be undergone for Models I to IV. To sum up the results for all the five models, $\lambda_i$ is ($i=I,II,III,IV,V$):
\begin{eqnarray}\label{VirialModels}
\mbox{ Model I} : \lambda_{I}=\frac{1-6\alpha_c}{2+3\alpha_c+3\alpha_x/R}\, ,  \nonumber \\ \nonumber \\
\mbox{ Model II} : \lambda_{II}=\frac{1-\frac{6\xi_1}{R+1}}{2+\frac{3\xi_1}{R+1}}\, ,  \nonumber \\ \nonumber \\
\mbox{ Model III} : \lambda_{III}=\frac{1}{2+\frac{3\xi_2}{R(R+1)}}\, ,  \nonumber \\ \nonumber \\
\mbox{ Model IV} : \lambda_{IV}=\frac{1-\frac{6R\xi_3}{R+1}}{2+\frac{3R\xi_3}{R+1}}\, ,  \nonumber \\ \nonumber \\
\mbox{ Model V} : \lambda_V=\frac{H-6\Gamma_c}{2H+3\Gamma_c+3\Gamma_x/R}\, .
\end{eqnarray}
Constant of the EOS, $w_j$, has a similar behavior for cold dark matter (CDM) and baryonic matter, that is $w_m=w_c=0$. Thus, Poisson equation or Eq.~(\ref{poisson}) leads to the same potential energy for both CDM and baryonic matter. It is very common to assume that baryons can merely interact with dark sector via gravitational field. In this case, which we call "First Possibility", Eq.~(\ref{virial2}) results in:
\begin{equation} \label{virial3} 
K=K_c+K_b=- \lambda_i U_G \, .
\end{equation}
Notwithstanding such a simple assumption, interaction between CDM and baryons might be considered a bit more intricate. Although both CDM and baryonic matter have the same potential function, they may interact separately, solely with their own type of matter. Given the circumstances, which we name "Second Possibility", Eq.~(\ref{virial2}) gives
\begin{equation} \label{virial4} 
K=-(\lambda_i \frac{\Omega_c}{\Omega_c+\Omega_b}+\frac{1}{2} \frac{\Omega_b}{\Omega_c+\Omega_b})U_G \, 
\end{equation}
where $\Omega$ is relevant density parameter for each element of matter. In order to brief calculations, we introduce parameter $\lambda_i'$ and write the last equation as
\begin{equation} \label{virial6}
\lambda_i'=\lambda_i \frac{\Omega_c}{\Omega_c+\Omega_b}+\frac{1}{2} \frac{\Omega_b}{\Omega_c+\Omega_b}\, ,
\end{equation}
\begin{equation} \label{virial5}
K=- \lambda_i' U_G\, .
\end{equation}
Eqs. (\ref{virial3}) and (\ref{virial5}) are the substitutes for the classical virial condition in dynamical equilibrium with respect to interaction between DE and DM (considering the First or the Second Possibilities). It is apparent that these equations with $\alpha_j=\xi_1=\xi_2=\xi_3=\Gamma_j=0$ reduce to the familiar $K=-\frac{1}{2}U$ in non-interacting models. \\

Note that the assumption of homogeneous distribution of DE in Eq.~(\ref{virial}) would be denied by non-standard models of DE. As an example, detecting fewer clusters than the prediction of the primary CMB anisotropies via the Sunyaev-Zel'dovich effect by Planck satellite \cite{Ade:2015fva} has given rise to the idea of clustering DE. In this regime, DE contributes to clustering, and hence, we cannot omit DE terms in Eq.~(\ref{Layzer-Irvine}) whereby virial theorem changes to a more intricate form (see \cite{Batista:2017lwf} and \cite{Chang:2017vhs} to find out how clustering DE model alters characteristics of virialized haloes). In this work, we consider the common standard DE and postpone more investigations on modified virial theorem with respect to DE with negligible sound speeds to future studies. \\

\section{Mass Temperature Relation of Galaxy Clusters}
%

The primary approach to form mass-temperature relation is combining the virial theorem with conservation of energy, which brings about $M\propto T^{\zeta}$. While the power-law index appears to be $\zeta=\frac{3}{2}$ in most masses, a "break" is predicted in a myriad of observations and simulations at low masses, which gives rise to $\zeta>\frac{3}{2}$ in this particular range. The physics behind this behavior has been under study for a while; \cite{Muanwong:2001fy} attributed it to the cooling process, and the heating process is stated in \cite{Bialek} to be the rationale for this "break", to name but a few. In order to reconstruct theories concerning this "break", Afshordi \& Cen have attributed it to the nonsphericity of the initial protoclusters in \cite{Afshordi:2001ze}, and Del Popolo has taken the angular momentum acquisition by protoclusters into account in \cite{Popolo}. Nevertheless, more recent studies, embracing \cite{Stanek} and \cite{Planelles}, revealed that there is no evidence of double slope in M-T relation. However, the existence of this "break" is still under discussion.\\ 

We try to take a look at three different methods which have been provided by Afshordi \& Cen and Del Popolo to reconstruct mass-temperature relation in galaxy clusters, considering the modified virial theorem for interacting dark matter and dark energy. The double slope in mass-temperature relation is not our principal focus and we neglect it, although there will be some mentions to that. \\

\subsection{Derivation of Mass-Temperature Relation}
In this section, we develop the method used by Afshordi \& Cen in \cite{Afshordi:2001ze} to rebuild M-T relation in galaxy clusters for interacting models. They begin with a definition of the kinetic and potential energies and pursue calculations by using velocity as a function of the gravitational potential in the perturbation theory, Poisson equation, and Gauss's theorem to obtain the initial energy of a protocluster (i.e. $E_{ta}$ or the total energy of that at turnaround radius $r_{ta}$). Since up to this point there is no indication of interacting dark sector, we avoid repeating calculations, and we just mention the outcome obtained in \cite{Afshordi:2001ze}:
 \begin{equation}\label{E_ta}
 E_{ta} = -\frac{10\pi G}{3} \rho_{ta}^2 r_{ta}^5 B \, .
 \end{equation}
Here, $B$ is defined as
\begin{equation}\label{B}
B \equiv \int_0^1\tilde{\delta}_{ta}(\tilde{r})
 (1-\tilde{r}^2)d^3\tilde{r}\, ,
\end{equation}
where $\tilde{r}\equiv\frac{r}{r_{ta}}$, $\tilde{\delta}_{ta} \equiv \delta_{ta} + \frac{3}{5}(\Omega_{ta}-1)$, and $\Omega_{ta}$ and $\delta_{ta}$ are density parameter and density contrast at turnaround time, respectively. \\

Taking a surface pressure term into account (which is exerted at the boundary of the cluster), virial condition gives
\begin{equation}\label{P_ext}
K_{vir}+E_{vir}=(1-2\lambda_i)U_{vir}+3P_{ext}V \, .
\end{equation}
There is the point where the impact of interacting dark sector emerges. Here, $P_{ext}$ denotes the pressure on the outer boundary of the virialized cluster, and $V$ stands for the volume. It is clear that the last equation could reduce to the classical equation (used by Afshordi \& Cen), if $\lambda_i=\frac{1}{2}$. Another equation for surface pressure is expressed by
\begin{equation}\label{P_ext_U}
3P_{ext}V = -\nu U_{vir}\, ,
\end{equation}
where the parameter $\nu$ is a coefficient constant to indicate the considered correlation between exerted pressure and the potential energy. Combining two preceding equations gives
\begin{equation}\label{K_E_U}
K_{vir}+E_{vir}=(1-2\lambda_i-\nu)U_{vir}\, .
\end{equation}
The surface pressure term also alters the relation between kinetic and potential energy after virialization to
\begin{equation}\label{K_U}
K_{vir}=-\frac{2\lambda_i+\nu}{2}U_{vir}\, .
\end{equation}
Inserting $U_{vir}$ from Eq.~(\ref{K_U}) into Eq.~(\ref{K_E_U}) leads to
\begin{equation}\label{K_E}
-\frac{2\lambda_i+\nu}{2-2\lambda_i-\nu}E_{vir}=K_{vir}\, .
\end{equation}
Then, the kinetic energy of the cluster can be separated into fully ionized baryonic gas and DM as
\begin{equation}\label{Kinetik_b_DM}
K_{vir} = \frac{3}{2} M_{c} \sigma_v^2+\frac{3M_{b} k_B T}{2\mu m_p}\, ,
\end{equation}
where $\sigma_v$ stands for the mass-weighted mean velocity dispersion of DM particles in one dimension, $M_{b}$ is the total baryonic mass, $k_B$ is the Boltzmann constant, T is temperature, $\mu=0.59$ is mean molecular weight and $m_p$ represents the proton mass. To simplify the previous equation, $\tilde{\beta}_{spec}$ is defined as
\begin{equation}\label{tilde_Beta}
\tilde{\beta}_{spec} = \beta_{spec} [1 + (f \beta_{spec}^{-1}-1)\frac{\Omega_b}{\Omega_b+\Omega_c}]\, .
\end{equation}
Here, $f$ is the fraction of baryonic matter in hot gas and $\beta_{spec} \equiv \sigma_v^2/(k_BT/\mu m_p)$. This definition assists to obtain from Eq.~(\ref{Kinetik_b_DM}):
\begin{equation}\label{kinetik_main}
K_{vir} = \frac{3\tilde{\beta}_{spec} M k_B T}{2\mu m_p}\, .
\end{equation}
Now, using Eqs. (\ref{E_ta}) and (\ref{kinetik_main}) in Eq. (\ref{K_E}), with respect to conservation of energy ($E_{ta}=E_{vir}$), we find:
\begin{equation}\label{kT}
k_B T = \frac{5 \mu m_p}{8 \pi \tilde{\beta}_{spec}} (\frac{2\lambda_i+\nu}{2-2\lambda_i-\nu}) H_{ta}^2 r_{ta}^2 B\, .
\end{equation}
In order to find an expression for $H_{ta}^2 r_{ta}^2$, parameter $e$ is defined to be the energy of a test particle with unit mass at $r_{ta}$, therefore, we can write it as
\begin{equation}\label{e}
e = \frac{\mathbf{v}^2_{ta}}{2}-\frac{GM}{r_{ta}}\, .
\end{equation}
We also have collapse time (or dynamical time scale) as
\begin{equation}\label{t}
t = \frac{2\pi G M}{(-2e)^{\frac{3}{2}}}\, .
\end{equation}
With the assumption that this time is approximately equal to the required time for virialization, and using the Friedmann equations, one can  obtain
\begin{equation}\label{HR}
-2e = \frac{5}{4 \pi}H_{ta}^2 r_{ta}^2 A = (\frac{2\pi G M}{t})^{\frac{2}{3}}\, ,
\end{equation}
\begin{equation}\label{A}
A \equiv \int_0^1 \tilde{\delta}_i(\tilde{r}) d^3\tilde{r} = \frac{2}{5}(\frac{3\pi^4}{t^2 G \rho_{ta}})^{\frac{1}{3}}\, .
\end{equation}
Using last two equations together with Eq. (\ref{kT}), the mass-temperature relation can be obtained
\begin{equation}\label{MT}
k_BT = (\frac{\mu m_p}{2\tilde{\beta}_{spec}})(\frac{2\lambda_i+\nu}{2-2\lambda_i-\nu})(\frac{2\pi GM}{t})^{\frac{2}{3}}(\frac{B}{A})\, .
\end{equation}
By inserting numerical values, this relation can be written as
\begin{equation}\label{MTQ}
k_BT = (6.62 keV) \tilde{Q} (\frac{M}{10^{15} h^{-1} M_{\odot}})^{2/3}\, ,
\end{equation}
where the dimensionless factor $\tilde{Q}$ is defined:
\begin{equation}\label{Q}
\tilde{Q} \equiv (\frac{\tilde{\beta}_{spec}}{0.9})^{-1}(\frac{2\lambda_i+\nu}{2-2\lambda_i-\nu})(\frac{B}{A}) (Ht)^{-2/3}\, .
\end{equation}
Eq. (\ref{MTQ}) is the mass-temperature relation in galaxy clusters, regarding interaction between DE and DM. It is noticeable that the effect of interacting dark sector is merely appeared in factor $\tilde{Q}$. Afshordi \& Cen have extensively discussed this factor in \cite{Afshordi:2001ze}. Overall, $\tilde{\beta}_{spec}$ is a function of the ratio of the kinetic energy per unit mass of DM to the thermal energy of gas particles ($\beta_{spec}$), the fraction of baryonic matter in hot gas ($f$) and the ratio of baryonic matter to DM in the sphere. According to different simulations and observations, these three parameters vary slightly whereby the final value for $\tilde{\beta}_{spec}$ does not face dramatic changes and is close to 0.9, hence we fix it by this figure in our calculations. The second variable, $\nu$, depends on density profile $f(\omega)$ and concentration parameter $c$, which is given by
\begin{equation}\label{nu}
\nu(c,f(\omega)) \equiv -\frac{3P_{ext}V}{U}= \frac{c^3\int_c^{\infty} f(\omega)g(\omega)\omega^{-2}d\omega}{\int_0^c f(\omega) g(\omega) \omega d\omega}\, ,
\end{equation}
where: 
 \begin{equation}\label{g}
g(\omega) = \int_0^\omega f(\omega) \omega^2 d\omega \, .
 \end{equation}
For density profile, we may choose NFW profile as: 
\begin{equation}\label{f}
f_{NFW}(\omega)=\frac{1}{(\omega)\,(1+\omega)^{2}}\, ,
\end{equation}
where $\omega=\frac{r}{r_s}$ and $r_s$ is the scale radius given in  \cite{Lokas:2000mu}. This profile is proposed by Navarro, Frenk and White and has been widely used and studied in literature. However, there have been some objections to that, as some recent observations have revealed a cored density profile in the inner region of the haloes. Several density profiles have been proposed to include the cored central region, including Burkert profile \cite{Burkert:1995yz}, which is expressed by
\begin{equation}\label{Moore}
f_{Burkert}(\omega)=\frac{1}{(1+\omega)\,(1+\omega^2)}\, .
\end{equation}
Clearly, considering each of these profiles may affect M-T relation, as well as the other properties of clusters. \\

Concentration parameter $c$ is defined as the ratio of virial radius to scale radius, that is $\frac{r_{vir}}{r_s}$. The density profile is exclusively described by $c$. In case there is not any observational data, the following relation (from \cite{Maccio:2008pcd}) may be used to find the value of the concentration parameter:
\begin{equation}\label{concentration}
c=8.3(\frac{M_{200}}{10^{12}M_{\odot}})^{-0.104}\, ,
\end{equation}
where $M_{200}$ is the mass enclosed by the radius in which the average density is 200 times the critical density of the universe. Meanwhile, mass-concentration relation has extensively been under study and it would have minuscule differences in various works, such as \cite{Bhattacharya}. \\

In Eq. (\ref{Q}), parameter $(\frac{B}{A})$ plays the prominent role in the "break" of mass-temperature relation in low masses. In spite of the fact that both $A$ and $B$ are proportional to scale factor, $\frac{A}{B}$ remains constant. Considering an initial density profile with multiple peaks (rather than a homogeneous distribution of density, or a profile with one central peak), Afshordi  \& Cen obtain
\begin{equation}\label{AtoB}
<\frac{B}{A}> = \frac{4(1-n)}{(n-5)(n-2)}[1-\frac{n(n+3)}{10(1-n)}(1-\Omega_c-\Omega_b-\Omega_{\Lambda}) (\frac{Ht}{\pi(\Omega_c+\Omega_b)})^{\frac{2}{3}}]\, ,
\end{equation}
where $n$ is the index of the density power spectrum. Choosing an initial density profile with multiple peaks would be more comprehensive and rational because, in hierarchical structure formation models, mass gradually accumulates in several regions of the initial cluster and not solely around the center. Taking nonsphericity in the geometry of the collapsing protocluster into account, which has a notable sign in low masses, Afshordi \& Cen write some equations for dispersion of factor $\frac{A}{B}$, or $\frac{\Delta B}{A}$. It reveals more dispersion in low masses and consequently, leads to a so-called "break" in M-T relation. However, as we have mentioned before, not only is there no agreement on the existence of this double slope, but there is also no sign of interacting dark sector in this parameter, thus, we neglect it for our study. \\

Furthermore, another parameter is introduced in \cite{Afshordi:2001ze} as:
\begin{equation}\label{parameter_y}
y=\frac{B}{A(Ht)^{\frac{2}{3}}}\, .
\end{equation}
This definition changes Eq. (\ref{Q}) to a more straightforward form. It can be written as a function of density profile and concentration parameter
\begin{equation}\label{y_relation}
 y(c,f) = \frac{\Delta^{1/3}(2-2\lambda_i-\nu)c \int_0^c f(\omega) g(\omega) \omega d\omega}{3\pi^{2/3} g^2(c)}\, ,
\end{equation}
where $\Delta$ is the overdensity of the sphere and for a virialized cluster is somewhere in the region of $\Delta=200$, meaning that the cluster has an average density of 200 times as much as critical density of the universe. Last relation is driven in \cite{Afshordi:2001ze} regarding virial theorem and the definition of $\nu$; meanwhile, owing to modification of virial theorem, the factor $(1-\nu)$ has changed to $(2-2\lambda_i-\nu)$ for interacting models. \\

Both the mass and temperature of a cluster have to be positive to result in a genuine outcome. Combining this principle with Eq. (\ref{Q}) shows a constraint on the possible values for $\lambda_i$. As all contributors in Eq. (\ref{Q}) are positive quantities, the ratio $(\frac{2\lambda_i+\nu}{2-2\lambda_i-\nu})$ should be positive. As a result, we should whether have
\begin{equation}\label{constraint}
-\frac{\nu}{2}<\lambda_i<\frac{2-\nu}{2} \, ,
\end{equation}
or
\begin{equation}\label{constraint_2}
\frac{2-\nu}{2}<\lambda_i<-\frac{\nu}{2} \, .
\end{equation}
Due to the fact that $\nu$ is always a positive parameter, Eq. (\ref{constraint_2}) necessitates a negative $\lambda_i$. Taking Eq. (\ref{virial3}) into account, a negative $\lambda_i$ does not have any physical meaning; thus, just Eq. (\ref{constraint}) could be acceptable as a criterion for the value of $\lambda_i$, and its more accurate form is
\begin{equation}\label{constraint_main}
0<\lambda_i<\frac{2-\nu}{2} \, .
\end{equation}
Note that Eq. (\ref{Q}) is derived for our "First Possibility". It is self-evident that by replacing "$\lambda_i$" with "$\lambda_i'$", we would also be able to study the "Second Possibility".\\

\subsection{Reforming the Top-Hat Model}
In order to form the "break" in M-T relation, Del Popolo takes angular momentum acquisition of the collapsing protoclusters into consideration in \cite{Popolo}, and later, reinforces this method by adding another term for dynamical friction in \cite{DelPopolo:2019oxn}. The angular momentum is acquired by interacting with neighboring protoclusters. Del Popolo suggests two approaches to formulate M-T relation. The first approach is based upon the development of the top-hat model and we investigate it in this section, with an additional assumption of the interacting dark sector.\\

To start this method, an ensemble of gravitationally growing mass concentrations is assumed and then with the assistance of the Liouville's Theorem, Del Popolo obtains the radial acceleration of a particle as
\begin{equation}\label{eq:coll}
\frac{{\rm d}v_r}{{\rm d}t} = -\frac{GM}{r^2} + \frac{L^2(r)}{M^{2}r^3} + \frac{\Lambda}{3}r -\eta\frac{{\rm d}r}{{\rm d}t}\, ,
\end{equation}
where $\eta$ is the dynamical friction coefficient and $L(r)$ denotes the acquired angular momentum in radius $r$ from the center of the cluster. $L(r)$ has a very complicated relation which can be found in \cite{R88a} and \cite{Popolo:1998fz}. Integrating the previous equation leads to
\begin{equation}\label{eq:coll1}
\frac{1}{2}\left(\frac{{\rm d}r}{{\rm d}t}\right)^{2} = \frac{GM}{r} + \int_0^r \frac{L^{2}}{M^{2}r^{3}}\mathrm{d}r + \frac{\Lambda}{6}r^{2} - \int_0^r \eta\frac{{\rm d}r}{{\rm d}t} + \epsilon \, ,
\end{equation}
Here, $\epsilon$ is the specific binding energy of the shell and can be determined by condition of $\frac{{\rm d}r}{{\rm d}t}=0$ at $r_{ta}$. The preceding equation represents four forms of potential energy; using them in the modified virial condition for interacting dark sector, we have
\begin{equation}\label{eq:virial}
 \langle K \rangle = -\lambda_i \langle U_{\rm G} \rangle - \langle U_{\rm L} \rangle + \langle U_{\Lambda}\rangle + \langle U_{\eta} \rangle\ \, .
\end{equation}
Here, $\langle \rangle$ indicates time averaged value of any quantity. By using Eq.~(\ref{P_ext_U}) and (\ref{K_E_U}) in the previous equation, we get
\begin{equation}\label{modified_virial}
\langle K \rangle = (2\lambda_i+\nu) (-\frac{1}{2} \langle U_{\rm G} \rangle - \langle U_{\rm L} \rangle + \langle U_{\Lambda}\rangle + \langle U_{\eta} \rangle)\, .
\end{equation}
Defining $r_{eff}$ as the time averaged radius of mass shell, Eq. (\ref{modified_virial}) can be written as
\begin{eqnarray}\label{eq:virial20}
& \langle K \rangle=-\left( \frac{2\lambda_i+\nu}{2}\right) U_{\rm G}\left[ 1+2\frac{U_{\rm L}}{U_{\rm G}}-2\frac{U_{\rm \Lambda }}{U_{\rm G}}-2\frac{U_{\rm \eta }}{U_{\rm G}} \right] \nonumber \\ \nonumber \\
& =\left( \frac{2\lambda_i+\nu}{2}\right) \frac{GM}{r_{\rm eff}}\left[ 1+2\frac{r_{\rm eff}}{GM^{3}}\int_{0}^{r_{\rm eff}}\frac{L^{2}(r)}{r^{3}}dr-\frac{\Lambda r_{\rm eff}^{3}}{3GM}-2\frac{r_{eff}}{GM}\int_{0}^{r_{eff}}\eta\frac{{\rm d}r}{{\rm d}t}\right] \, .
\end{eqnarray}
Ratio of $r_{eff}$ to $r_{ta}$ is defined by $\psi=\frac{r_{eff}}{r_{ta}}$; then we have
\begin{eqnarray}\label{psi}
M=4 \pi \rho_{\rm b} x^{3}_{1}/3 \, ,  \nonumber \\ \nonumber \\
\chi=r_{ta}/x_{1}\, ,  \nonumber \\ \nonumber \\
\Omega_0=\frac{8 \pi G \rho_{\rm b}}{3 H_0^2} \, ;
\end{eqnarray}
and as a result:
\begin{equation}\label{R_eff}
r_{eff}=\psi \chi \left( \frac{2GM}{\Omega _{0}H_{0}^{2}}\right)^{1/3}\, .
\end{equation}
Then, putting $\langle K \rangle$ from Eq. (\ref{kinetik_main}) into Eq. (\ref{eq:virial20}) results in the M-T relation as
\begin{eqnarray}\label{M-T_popolo}
\frac{k_{\rm B}T}{{\rm keV}} = 1.58\left(\lambda_i+\nu\right) \frac{\mu}{\beta_{spec}}\frac{1}{\psi\chi} \Omega_{\rm 0}^{1/3}\left(\frac{M}{10^{15}M_{\odot}h^{-1}}\right)^{2/3}(1+z_{\rm ta}) \nonumber \times \nonumber \\ \nonumber \\
\left[1+\left(\frac{32\pi}{3}\right)^{2/3} \psi\chi \rho_{\rm b,ta}^{2/3} \frac{1}{H_{0}^{2}\Omega_{\rm b,0}M^{8/3}(1+z_{\rm ta})}\right. \nonumber \phantom{\times \left[\right.}
\times \int_0^{r_{\rm eff}} \frac{L^{2}}{r^{3}}{\rm d}r - \frac{2}{3}\frac{\Lambda}{\Omega_{\rm b,0}H_0^{2}(1+z_{\rm ta})^3}\left(\psi\chi\right)^3 \nonumber\phantom{\times \left[\right.} \nonumber \\ \nonumber \\
-\frac{2^{10/3}}{3^{2/3}} \pi^{2/3} \left(\frac{\psi\chi}{\Omega_{\rm b,0} H_0^2}\right) \left(\frac{\rho_{\rm b,0}}{M}\right)^{2/3} \frac{1}{1+z_{\rm ta}} 
\times \phantom{\times \left[\right.}\left.\int \eta \frac{{\rm d}r}{{\rm d}t}{\rm d}r\right]\, .
\end{eqnarray}
Conservation of energy should be used in order to determine the value of $\psi$, or $r_{eff}$ as
\begin{equation}\label{conservation}
\langle E \rangle= \langle K \rangle+ \langle U_{\rm G} \rangle+\langle U_{\rm \Lambda} \rangle+\langle U_{\rm L} \rangle+\langle U_{\rm \eta} \rangle=U_{\rm G,ta}+U_{\rm \Lambda,ta}+U_{\rm L,ta}+U_{\rm \eta,ta}\, .
\end{equation}
Using Eq. (\ref{modified_virial}) in this equation, we find
\begin{equation}\label{conservation2}
\frac{-2\lambda_i-\nu+2}{2}\langle U_{\rm G}\rangle-(2\lambda_i+\nu-1)\langle U_{\rm L} \rangle+(2\lambda_i+\nu+1)(\langle U_{\rm \Lambda} \rangle+\langle U_{\rm \eta} \rangle)=U_{\rm G,ta}+U_{\rm \Lambda,ta}+U_{\rm L,ta}+U_{\rm \eta,ta}\, ,
\end{equation}
and with the aid of the method provided by  \cite{Lahav:1991wc} for the last equation, the cubic equation below is obtained
\begin{eqnarray}\label{lahav}
(-2\lambda-\nu+2)+\left(\chi\psi\right)^{3}\left(2\lambda_i+\nu+1\right)\Upsilon - \psi \left(2+\Upsilon\chi^3\right) \nonumber\\ \nonumber \\
-\frac{27}{32}\frac{\chi^{9}\psi}{\rho_{\rm ta}^{3}\pi^{3}Gr_{\rm ta}^{8}}\left[(2\lambda_i+\nu-1)\int_{0}^{r_{\rm eff}}\frac{L^{2}(r)}{r^{3}}\mathrm{d}r + 
          \int_{0}^{r_{\rm ta}}\frac{L^{2}(r)}{r^{3}}\mathrm{d}r \right.\nonumber \\ \nonumber \\
  -  \frac{16\pi^2}{9}(2\lambda_i+\nu+1)\rho_{\rm ta}^2 r_{\rm ta}^6 \times \nonumber \left.\left(
          \int_{0}^{r_{\rm eff}}\eta\frac{\mathrm{d}r}{\mathrm{d}t}\mathrm{d}r - 
          \frac{1}{2\lambda_i+\nu+1}\int_{0}^{r_{\rm ta}}\eta\frac{\mathrm{d}r}{\mathrm{d}t}\mathrm{d}r\right)\right]=0 \, ,
\end{eqnarray}
with
\begin{equation}\label{upsilon}
\Upsilon=\frac{\Lambda}{4 \pi G \rho_{\rm ta}}=
\frac{\Lambda r_{\rm ta}^3}{3 G M}= 
\frac{2 \Omega_{\Lambda}}{\Omega_0}
\left(\frac{\rho_{\rm ta}}{\rho_{\rm ta,b}}\right)^{-1}(1+z_{\rm ta})^{-3} \, .
\end{equation}
Then it is possible to find $\psi$, or $r_{eff}$ by solving the above equation. Note that M-T relation or Eq. (\ref{M-T_popolo}) can be expressed in terms of $r_{vir}$ as
\begin{eqnarray}\label{M-T_popolo_new}
 \frac{k_{\rm B}T}{\rm keV} = 0.94\left(2\lambda_i+\nu\right)\frac{\mu}{\beta_{spec}}\left(\frac{r_{\rm ta}}{r_{\rm vir}}\right)
                  \left(\frac{\rho_{\rm ta}}{\rho_{\rm b,ta}}\right)^{1/3}\Omega_{\rm 0}^{1/3}\left(\frac{M}{10^{15}M_{\odot}h^{-1}}\right)^{2/3}(1+z_{\rm ta})\nonumber\\ \nonumber \\
              \times \left[1+\frac{15r_{\rm vir}\rho_{\rm b,ta}}{\pi^{2}H_0^{2}\Omega_{\rm 0}
                  \rho_{\rm ta}^{3}r_{\rm ta}^{9}(1+z_{\rm ta})}\int_0^{r_{\rm vir}} \frac{L^2(r) \mathrm{d}r}{r^3}
                  \right. -\frac{2}{3}\frac{\Lambda}{H_0^{2}\Omega_{0}}\left(\frac{r_{\rm vir}}{r_{\rm ta}}\right)^{3}
                  \left(\frac{\rho_{\rm b,ta}}{\rho_{\rm ta}}\right) \frac{1}{(1+z_{\rm ta})^3}\nonumber\\ \nonumber \\
 -\frac{6^{1/3}}{\pi^{1/3}}r_{\rm vir}r_{\rm ta}
                  \left(\frac{\rho_{\rm b,ta}}{\rho_{\rm ta}}\right)^{1/3}
                  \left(\frac{\rho_{\rm b,0}}{M} \right)^{2/3}\frac{1}{1+z_{\rm ta}}\times \left.\frac{\lambda_0}{1-\mu(\delta)}\right]\, ,
\end{eqnarray}
where $\mu(\delta)$ and $\lambda_0$ are parameters related to dynamical friction and are given in  \cite{Colafrancesco:1994ne}.\\ 

The previous equation has obtained for the mass-temperature relation of galaxy clusters, considering the effects of angular momentum acquisition (in \cite{Popolo}), dynamical friction (in \cite{DelPopolo:2019oxn}) and eventually, the impact of interacting dark sector, in this paper. As can be seen, $\lambda_i$ plays a more profound role in this approach, in comparison with Afshordi \& Cen's method, owing to its contribution to both Eqs. (\ref{lahav}) and (\ref{M-T_popolo_new}). Similar to the preceding model, "$\lambda_i'$" could be substituted for "$\lambda_i$" to create the "Second Possibility" in all equations.\\ 

This model is based on the assumption of cluster formation with the evolution of a spherical top-hat density perturbation, and the "late-formation approximation". The latter approximation states that any cluster at redshift $z$ is just reached its virialization. Although it is a good assumption in some cases, including the critical case of $\Omega_0=1$ (where the cluster formation is rapid), it constructs impediments to other cosmological models. \\

\subsection{Continuous Formation Model}
After a discussion on limitations and disadvantages to the former model in  \cite{Popolo}, Del Popolo derives M-T relation concerning the continuous formation model, which had been used in  \cite{Voit:2000ie} before. In this model, cluster formation occurs gradually, instead of instantaneously. The effects of angular momentum and dynamical friction with respect to this approach have been studied in \cite{Popolo} and \cite{DelPopolo:2019oxn}, respectively. Now, we are going to study how interacting dark sector makes a difference in M-T relation in terms of this procedure.\\

By integrating Eq. (\ref{eq:coll}), Del Popolo obtains an expression for the ratio of the total energy of a virialized cluster to its mass or $\frac{E}{M}$. We avoid iterating calculations, so the result is
\begin{equation}\label{E_M}
\frac{E}{M}=\frac{3m}{10(m-1)}\left( \frac{2\pi G}{t_\Omega }\right) ^{\frac 23}M^{\frac 23}\left[\frac{1}{m} + \left(\frac{t_\Omega}{t}\right)^{2/3} + \frac{K(m,x)}{(M/M_0)^{8/3}} + \frac{\lambda_0}{1-\mu(\delta)} \right.\left.\quad +\frac{\Lambda\chi^3}{3H^2_0\Omega_{\rm b,0}}\right]\, ,
\end{equation}
where
\begin{eqnarray}\label{parameters}
& t_\Omega =\frac{\pi \Omega _0}{H_o\left( 1-\Omega _0-\Omega _\Lambda \right) ^{\frac 32}} \, , \nonumber\\ \nonumber \\
& K(m,x)=\left (m-1\right )F x {\it LerchPhi}(x,1,3m/5+1)-\left (m-1\right )F{\it LerchPhi}(x,1,3m/5) \, , \nonumber\\ \nonumber \\
& LerchPhi(x',y',z')=\sum_{n=0}^{\infty} \frac{x'^n}{(z'+n)^{y'}} \, , \nonumber\\ \nonumber \\
& F=\frac{2^{7/3} \pi^{2/3} \chi \rho_{\rm b}^{2/3}}{3^{2/3} H^2 \Omega} \int_0^r \frac{L^2(r) dr}{r^3} \, , \nonumber\\ \nonumber \\
& x=1+(\frac{t_{\Omega}}{t})^{2/3}\, .
\end{eqnarray}
Meanwhile, $M=M_0 x^{-{3m/5}}$ and $M_0$ is given in  \cite{Voit:2000ie}.\\ 

Combining Eqs. (\ref{E_M}) and (\ref{kinetik_main}) with the virial theorem results in
\begin{equation}\label{T_E_M}
 k_{\rm B}T = \frac{4}{3}\tilde{a}\frac{\mu m_{\rm p}}{2\beta_{spec}}\frac{E}{M}\, ,
\end{equation}
and afterwards
\begin{eqnarray}\label{continuous_M_T}
\frac{k_{\rm B}T}{\rm keV} = \frac{2}{5}\tilde{a}\frac{\mu m_{\rm p}}{2\beta_{spec}} 
 \frac{m}{m-1}\left(\frac{2\pi G}{t_\Omega}\right)^{2/3}M^{2/3}
\times\nonumber\\ \nonumber \\
\left[\frac{1}{m} + \left(\frac{t_\Omega}{t}\right)^{2/3} + \frac{K(m,x)}{(M/M_0)^{8/3}} + \frac{\lambda_0}{1-\mu(\delta)} \right.\left.\quad +\frac{\Lambda\chi^3}{3H^2_0\Omega_{\rm b,0}}\right]\, .
\end{eqnarray}
Here, the parameter $\tilde{a}$ is the ratio of the kinetic to total energy of the cluster, and according to Eq. (\ref{K_E}) we have
\begin{equation}\label{a_tilde}
\tilde{a}=\frac{2\lambda_i+\nu}{2-2\lambda_i-\nu}\, .
\end{equation}
Thus, we can see that the trace of interacting dark sector emerges in a factor in M-T relation. Likewise the previous procedures, putting "$\lambda_i'$" instead of "$\lambda_i$" gives the equation for the "Second Possibility". \\

 \section{Results and Discussion}
 
We use five different sets of observational data to determine constants of the interacting dark sector for Models I to V, with the aid of M-T relation (Eq. (\ref{MTQ})). These observational data sets are provided in \cite{Horner:1999wf}, \cite{Finoguenov:2000qb}, \cite{Vikhlinin:2005mp}, \cite{Vikhlinin:2008cd} and \cite{Lovisari:2014pka}. The first set provides details of mass and temperature for 32 clusters (hereafter Obs. 1999). The second source of data is used by Afshordi \& Cen in \cite{Afshordi:2001ze} and consists of 39 clusters (hereafter Obs. 2001). The third data set includes Chandra's observations for 10 low-redshift clusters (hereafter Obs. 2006) and details of 49 low-redshift clusters from Chandra are collected in the fourth data set (hereafter Obs. 2009). Finally, the last resource comprises 20 clusters from XMM-Newton observations (hereafter Obs. 2015). \\

Measurements of temperature are generally based on X-ray observations, hence the temperatures given in the mentioned catalogs are X-ray temperature and could be different than density-weighted temperature in Eq. (\ref{MTQ}), which is averaged over the whole cluster. The reason lies within the fact that X-ray temperature ($T_X$) is exclusively measured over the central brighter portion of the cluster. To convert X-ray temperature to $T$ in Eq. (\ref{MTQ}), we use the relation below from  \cite{temperature} : 
\begin{equation}\label{T_convert}
 T=T_X[1+(0.22\pm 0.05)\log_{10}T_X(keV)-(0.11\pm 0.03)] \, .
\end{equation}
As it has been mentioned before, it is prevalent to consider the overdensity of the virialized clusters to be about 200 times the critical density of the universe. Therefore, $M_{200}$ is considered to be the cluster mass after virialization. The masses given in Obs. 1999 to 2015 have been obtained with respect to different methods and none of them incorporates $M_{200}$. In order to convert these masses to $M_{200}$ (e.g. $M_{500}$ to $M_{200}$), we use the relation $M_\delta \propto \delta^{-0.266}$ from  \cite{Horner}, where $\delta=\frac{M(<r)}{\frac{4}{3}\pi \rho_c r^3}$. This relation have been obtained via fitting the relation of $M\propto T^{\frac{3}{2}}$ to simulation data, regarding different values of $\delta$. As our calculations revealed, considering interaction between DM and DE has no impact on density profile and it only affects the factor of M-T relation. Therefore, this relation can be used for mass conversion. \\

Our aim is to fit observational data between $M_{200}$ and $T$ to the relation of $M \propto T^{\frac{3}{2}}$, in order to find the matched value of $\lambda_i$ in coefficient factor for each fit and each model and then determine the interacting constants. In addition, some values for constants of the interacting dark sector have been recently proposed in  \cite{vonMarttens:2018iav} for Models II, III, IV, and two special cases of Model I, based on various observations. Observations related to Type-Ia Supernovae (SNe Ia), the present value of the Hubble parameter ($H_0$), cosmic chronometers (CC), baryon acoustic oscillations (BAO), and the Planck measurements of the CMB temperature anisotropy (Planck TT) are the five types of observation which have been made up constraints in \cite{vonMarttens:2018iav} to find constants of the interacting dark sector. We also use those proposed values in M-T relation to make comparison among outcomes and observational data sets for mass and temperature.\\ 

\begin{figure}[h]
 	\centering
 	\includegraphics [width=0.63 \linewidth] {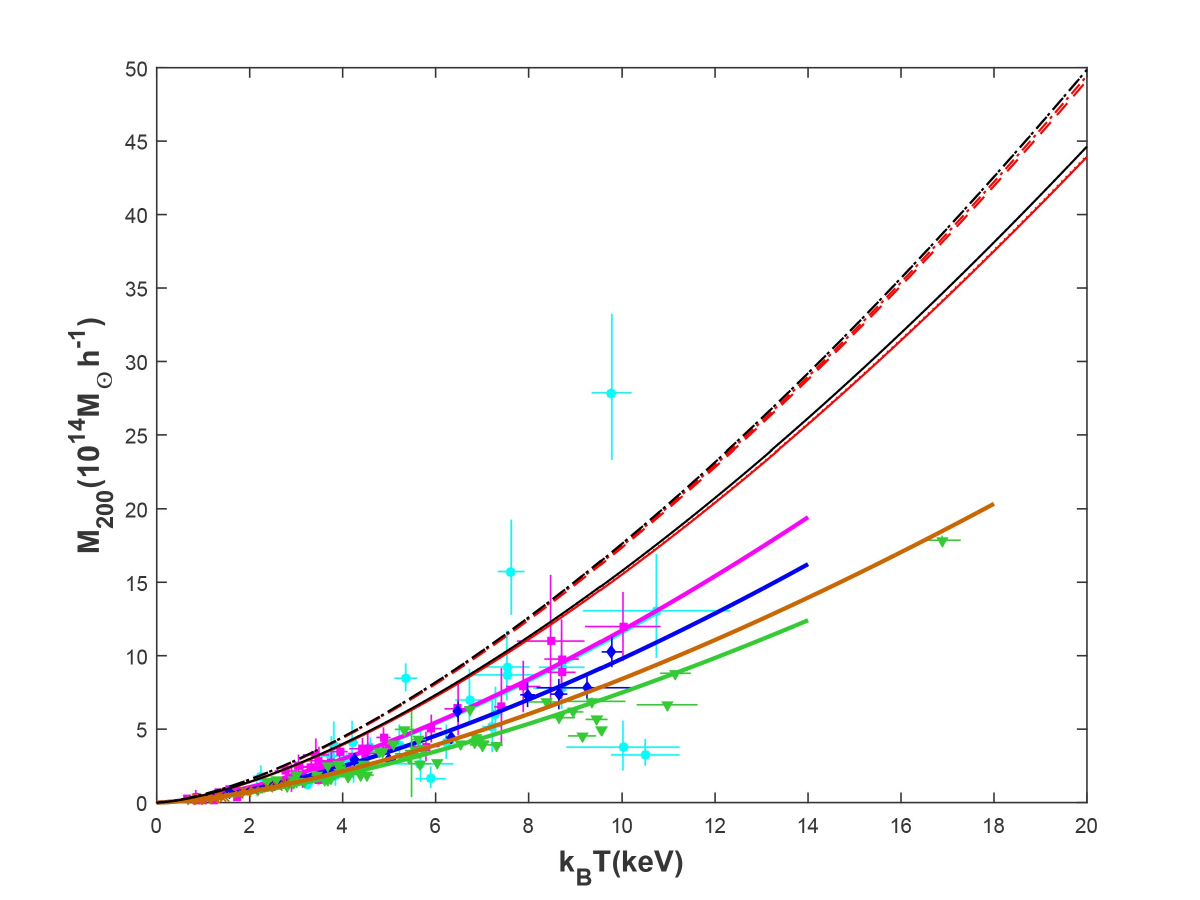}
 	\caption{The behavior of the mass-temperature relation in interacting Model I, in special case of $\alpha_x=0$. Red lines indicate the outcome of "$SNe \, Ia + H_0 + CC + BAO$" observations for $\alpha_c$ and black lines display the prediction related to "$Planck \, TT$" observations for this parameter. The other five colors denote five observational data sets from 1999 to 2015 (Obs. 1999: cyan; Obs. 2001: magenta; Obs. 2006: blue; Obs. 2009: green; Obs. 2015: brown). Solid and dotted lines show the "First" and "Second" possibilities for NFW density profile, while dashed and dash-dot lines illustrate these two possibilities for Burkert profile, respectively.}
 	\label{figure_1}
 \end{figure}

Model I is expressed by two interacting constants, namely $\alpha_x$ and $\alpha_c$. Two specific and simple cases for this model are $\alpha_x=0$ and $\alpha_c=0$. Fig. (\ref{figure_1}) compares the mass-temperature relation under the assumption of $\alpha_x=0$ for Model I with observational data sets, based on the values obtained in  \cite{vonMarttens:2018iav}. According to  \cite{vonMarttens:2018iav}, observations of "$SNe \, Ia + H_0$" and "$SNe \, Ia + H_0 + CC$" result in $\alpha_c=-0.36$ and $\alpha_c=-0.092$, respectively. These two values are not consistent with the constraint of Eq. (\ref{constraint_main}) and give the unreal negative temperatures for given masses. The outcome of $\alpha_c=-0.0019$, which is obtained from "$SNe \, Ia + H_0 + CC + BAO$", is illustrated with red lines in Fig. (\ref{figure_1}). Likewise, the constraint of "$Planck \, TT$" has given $\alpha_c=-9.73\times10^{-5}$ and its result in M-T relation is shown with black lines. For both predictions, solid lines are related to the "First Possibility" of the NFW density profile, while dotted lines are attributed to the "Second Possibility" for the same density profile. The results of the Burkert density profile are presented by dashed lines (for the "First Possibility") and dash-dot lines (for the "Second Possibility"). Note that the differences between first and second possibilities are very subtle in this model whereby solid and dotted black lines are almost indistinguishable. Data sets and their fitted curves for Obs. 1999, 2001, 2006, 2009, and 2015 are demonstrated with colors cyan, magenta, blue, green, and brown, respectively (the fitted lines for Obs. 1999 and 2001 are virtually coincident). \\

We immediately infer that for the case of $\alpha_x=0$ in Model I, a negative $\alpha_c$ has to be very close to zero to not violate the constraint of Eq. (\ref{constraint_main}). However, these values are not consistent with any observational data set. \\

\begin{figure}[h]
 	\centering
 	\includegraphics [width=0.63 \linewidth] {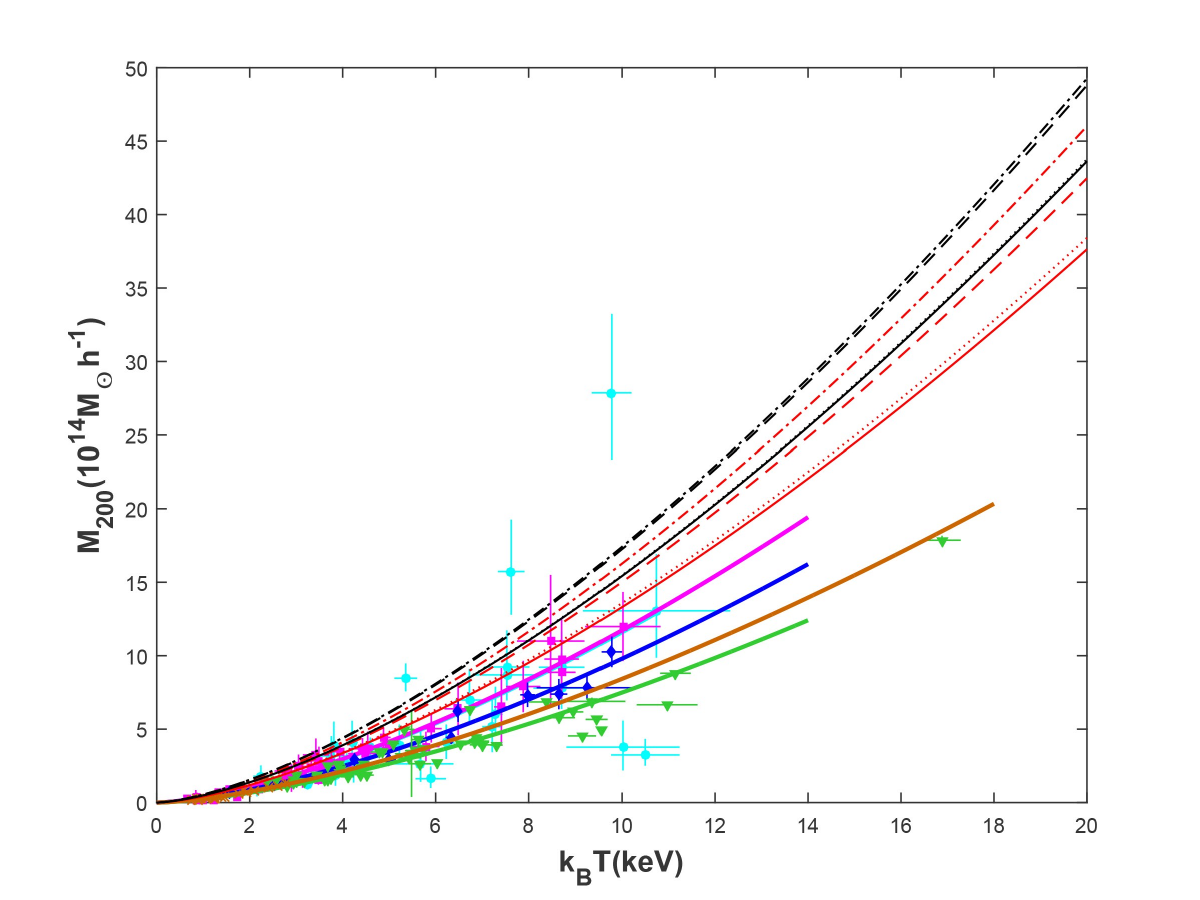}
 	\caption{Comparison between observational data and the predictions of "$SNe \, Ia + H_0 + CC + BAO$" and "$Planck \, TT$" observations for the case of $\alpha_c=0$ in Model I. Colors and types of lines are chosen the same as Fig. (\ref{figure_1})}
 	\label{figure_2}
 \end{figure}

Fig. (\ref{figure_2}) indicates M-T relation for another special case for Model I, which is $\alpha_c=0$. Chosen colors and types of lines are the same as Fig. (\ref{figure_1}) and again, results of "$SNe \, Ia + H_0$" (with $\alpha_x=-0.26$) and "$SNe \, Ia + H_0 + CC$" (with $\alpha_c=-0.27$) violate the constraint of Eq. (\ref{constraint_main}) and consequently, cannot be presented. Whereas, observations of "$SNe \, Ia + H_0 + CC + BAO$" (with $\alpha_x=-0.037$) and "$Planck \, TT$" (with $\alpha_x=-0.0052$) are theoretically acceptable. Here, the difference between the first and second possibilities is easier to spot, in comparison with the former case. As it can be seen, these predicted values lead to higher masses than observational data, likewise the previous case.  \\

\begin{figure}[h]
 	\centering
 	\includegraphics [width=0.63 \linewidth] {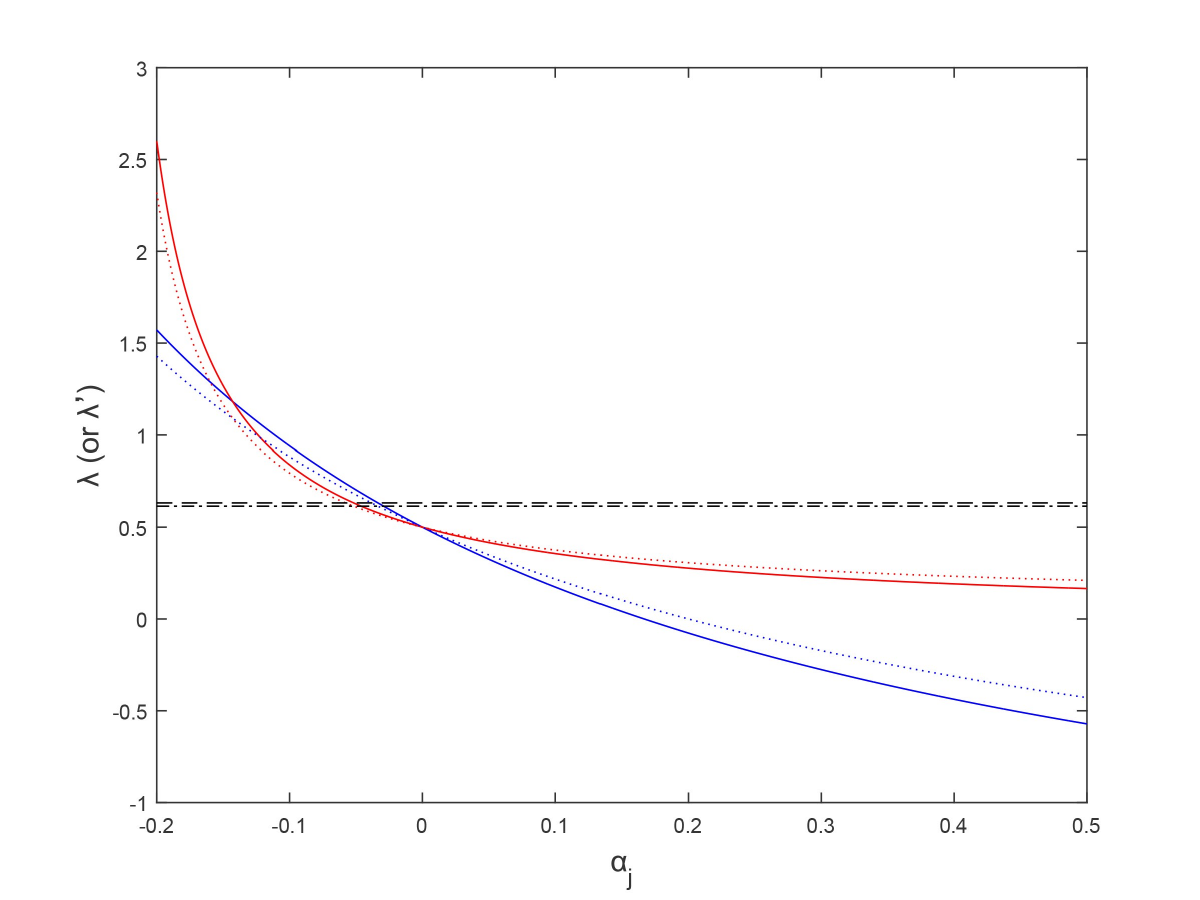}
 	\caption{The behavior of $\lambda_I$ (or $\lambda_I'$) as a function of interacting constant for two simple cases of Model I. The black dashed line represents the result of Obs. 1999 for the "First Possibility" and the dash-dot line shows this for the "Second Possibility". The blue lines are related to the case of $\alpha_x=0$ and the red lines indicate Model I with $\alpha_c=0$. Here, the solid lines describe the "First Possibility", while the dotted lines are attributed to the "Second Possibility".}
 	\label{Comparison Model I}
 \end{figure}

The value of $\lambda_i$ explains the ratio of kinetic to potential energy after virialization and plays the most vital role in our calculations. Fig. (\ref{Comparison Model I}) reveals how $\lambda$ and $\lambda'$ change as a function of $\alpha_c$ or $\alpha_x$, in two mentioned cases of Model I, which are more simple. The blue lines are related to Model I with $\alpha_x=0$ and the red lines describe the same model with $\alpha_c=0$. Therefore, the horizontal axis is attributed to $\alpha_c$ in the former case, and to $\alpha_x$ in the latter one. Moreover, solid lines are shown as the symbol of the "First Possibility", and the dotted lines denote the "Second Possibility", mutually. The black dashed line is drawn with respect to the obtained value of $\lambda_I$ for Obs. 1999 ("First Possibility"); and the dash-dot line shows the same value, but regarding the "Second Possibility". We do not display the outcomes of the other four observational data sets to avoid an overcrowded graph.\\

 For both situations of Model I, large negative values of $\alpha_j$ are too far away from the observational results. As the interacting constants are declining, both cases reach to observational outcomes just before the zero points. Although Model I with $\alpha_x=0$ almost keeps its slope for positive values, the case of $\alpha_c=0$ remains stable and would be rather comparable with observational results if $\lambda$ and $\lambda'$ were less than 0.5, even for higher values of $\alpha_x$. According to the definition of $Q$ for Model I, it means that if the transfer of energy from DE to DM primarily stemmed from the density of DE, different values for interacting constant would not lead to considerable changes in the virial condition. In other words, whether the protocluster consists of a dense region of DE or not, there would be merely negligible differences. However, it does not have great practical importance, since we initially assumed that the distribution of DE is unchanged through the interior and exterior of the collapsing sphere. On the contrary, if the energy transfer between DE and DM were mostly affected by the density of DM, the virial theorem would gradually change with interacting constant. \\

\begin{figure}[h]
 	\centering
 	\includegraphics [width=0.63 \linewidth] {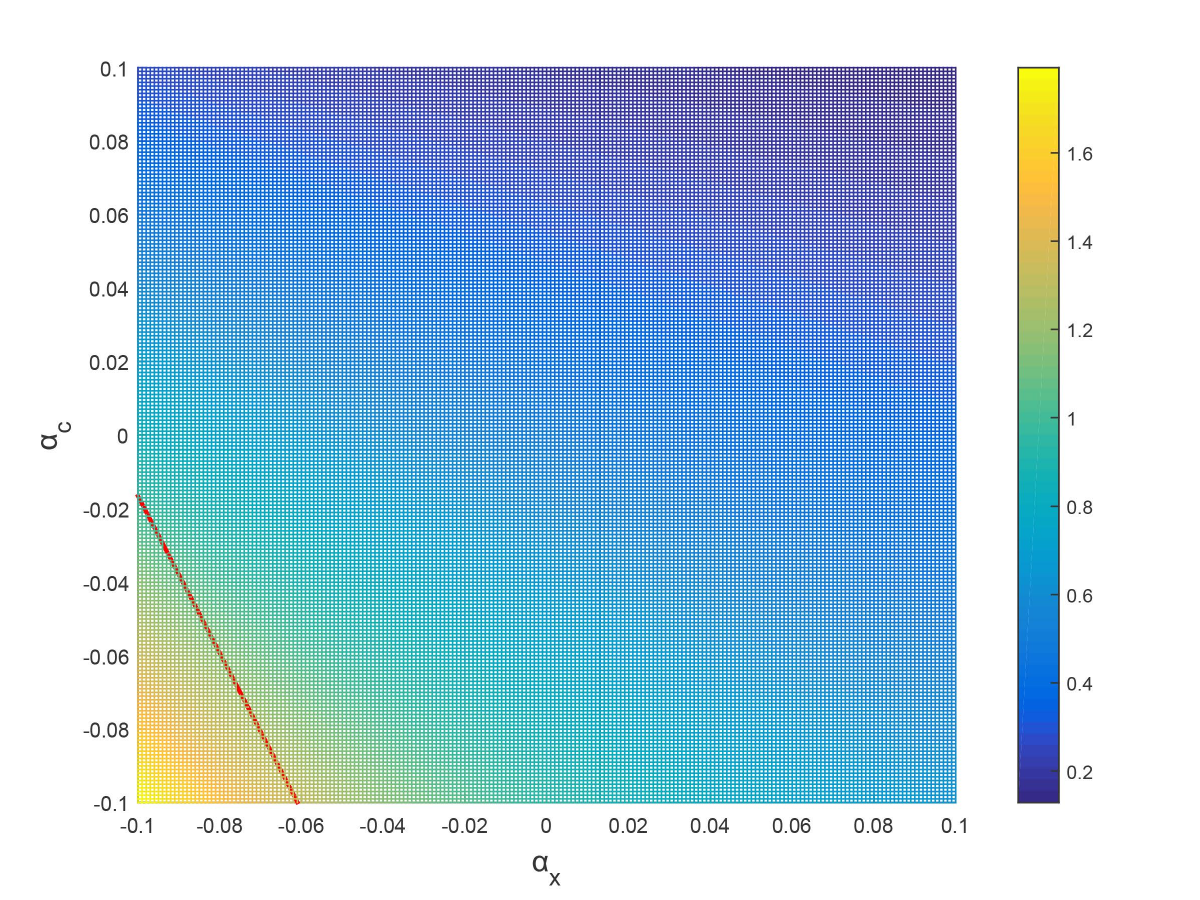}
 	\caption{Different combinations of $\alpha_c$ and $\alpha_x$ in the range between $-0.1$ to $0.1$ result in the value of $\lambda_I$ from just less than 0.2 to over 1.6, as it is illustrated in this figure. Colors stand for the given value of $\lambda_I$ for any given couple of $\alpha_c$ and $\alpha_x$, according to the guide strip in the right side. The red line specifies the obtained constraint, which confines real physical choices.}
 	\label{Model_I graph}
 \end{figure}

Description of Model I in general (without any zero constant) is more elaborate. Nonetheless, several constraints have been yet derived. For example,  \cite{CalderaCabral:2008bx} obtains four constraints between $\alpha_c$ and $\alpha_x$. In our study, Eq. (\ref{constraint_main}) gives rise to another constraint for these two parameters:
\begin{equation}\label{constraint_model_I}
0<\frac{1-6\alpha_c}{2+3\alpha_c+3\alpha_x/R}<\frac{2-\nu}{2} \, .
\end{equation}
Fig. (\ref{Model_I graph}) illustrates how different inputs of $\alpha_c$ and $\alpha_x$ give different amounts of $\lambda_I$, for a small range from $-0.1$ to $0.1$ as an example. Colors denote different values of $\lambda_I$ for each given $\alpha_c$ and $\alpha_x$. The red line also constrains acceptable choices for these two parameters, according to Eq. (\ref{constraint_model_I}). Here, we chose the value of $c=5$ for a typical cluster and used NFW density profile to calculate $\nu$. All the points in the left-bottom corner of the figure (below the red line) are unacceptable and have no physical meaning due to our recent constraint. In this specific region, which has been deliberately chosen to be close to non-interacting models, every couple with $\alpha_c=-\alpha_x$ gives approximately the same value for $\lambda_I$, while $\alpha_c=\alpha_x$ results in very different numbers. \\

\begin{figure}[h]
 	\centering
 	\includegraphics [width=0.63 \linewidth] {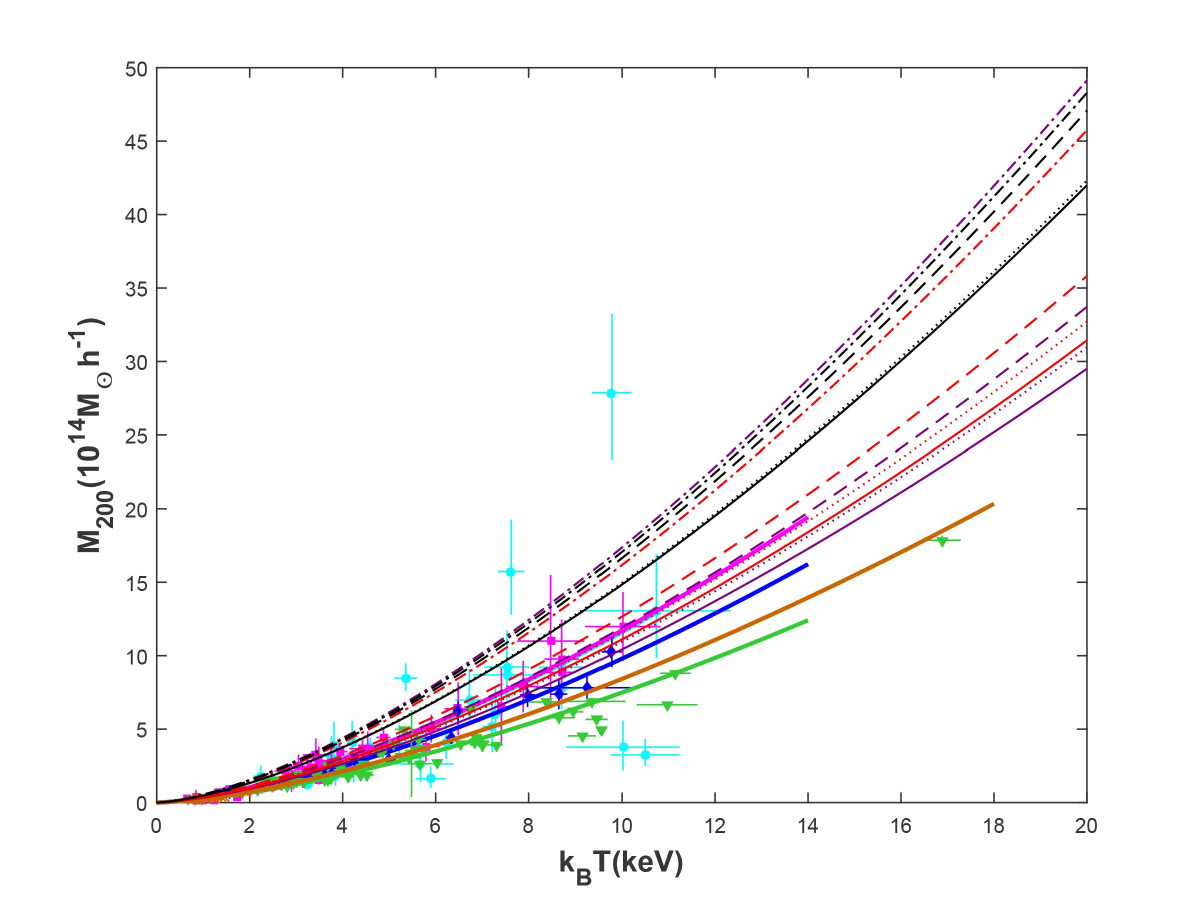}
 	\caption{The M-T diagram of galaxy clusters based on Model II. The features are identical to Figs. (\ref{figure_1}) and (\ref{figure_2}), except for the purple lines which are emerged because the predicted value from "$SNe \, Ia + H_0 + CC$" observations is allowable in this model.}
 	\label{figure_3}
 \end{figure}

For Models II, III and IV, there is only one interacting constant. For Model II, Fig. (\ref{figure_3}) makes a comparison between observational data and the outcome of obtained values for $\xi_1$ in  \cite{vonMarttens:2018iav}. Similar to the previous cases, the result of "$SNe \, Ia + H_0$", which has given $\xi_1=-0.53$, violates Eq. (\ref{constraint_main}) and leads to negative temperatures. Despite Model I, "$SNe \, Ia + H_0 + CC$" (with $\xi_1=-0.07$) results in an allowable prediction for M-T relation, which is represented with the purple lines in Fig. (\ref{figure_3}). The characteristics of the other lines are selected similar to Figs. (\ref{figure_1}) and (\ref{figure_2}); with $\xi_1=-0.06$ for "$SNe \, Ia + H_0 + CC + BAO$" and $\xi_1=-0.010$ for "$Planck \, TT$". In this model, predictions of "$SNe \, Ia + H_0 + CC$" and "$SNe \, Ia + H_0 + CC + BAO$" are close to some observational data. For example, "$SNe \, Ia + H_0 + CC + BAO$" result of the "Second Possibility" in NFW density profile and also the outcome of "$SNe \, Ia + H_0 + CC$" for the "First Possibility" in Burkert density profile are approximately in agreement with Obs. 1999 and Obs. 2001. \\

\begin{figure}[h]
 	\centering
 	\includegraphics [width=0.63 \linewidth] {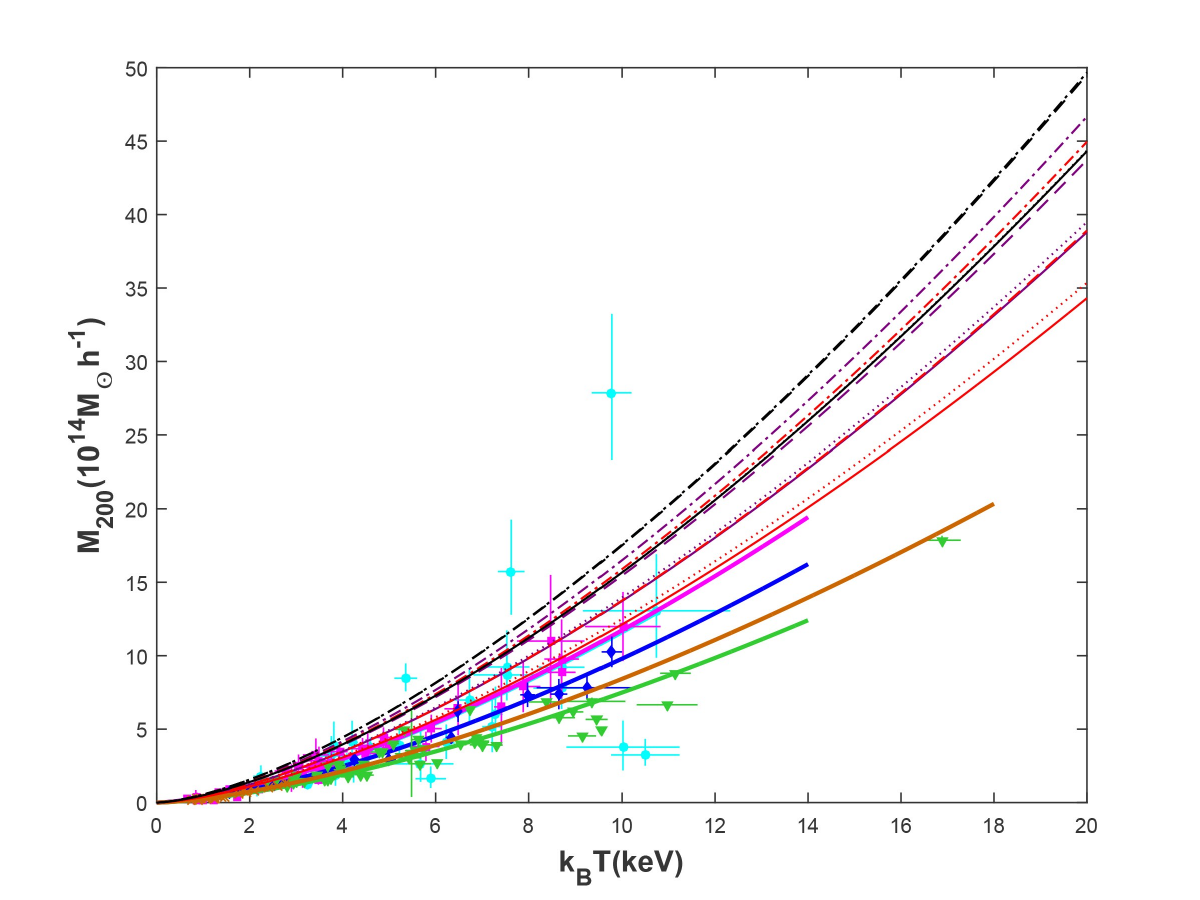}
 	\caption{The M-T diagram for Model III; all chosen colors and types of lines are analogous to Fig. (\ref{figure_3})}
 	\label{figure_4}
 \end{figure}

Similarly, Fig. (\ref{figure_4}) shows M-T relation for three allowable values of $\xi_2$ in Model III and compares them with fitted curves of the five observational data sets. Here, the values have been proposed as: $\xi_2=-0.40$ for "$SNe \, Ia + H_0$" (unacceptable), $\xi_2=-0.04$ for "$SNe \, Ia + H_0 + CC$" (purple lines), $\xi_2=-0.08$ for "$SNe \, Ia + H_0 + CC + BAO$" (red lines) and $\xi_2=-0.0024$ for "$Planck \, TT$" (black lines). It is clear that merely, the results of "$SNe \, Ia + H_0 + CC + BAO$" for the NFW density profile are almost close to Obs. 1999 and Obs. 2001 and again, the other predictions show higher masses than data sets. \\

\begin{figure}[h]
 	\centering
 	\includegraphics [width=0.63 \linewidth] {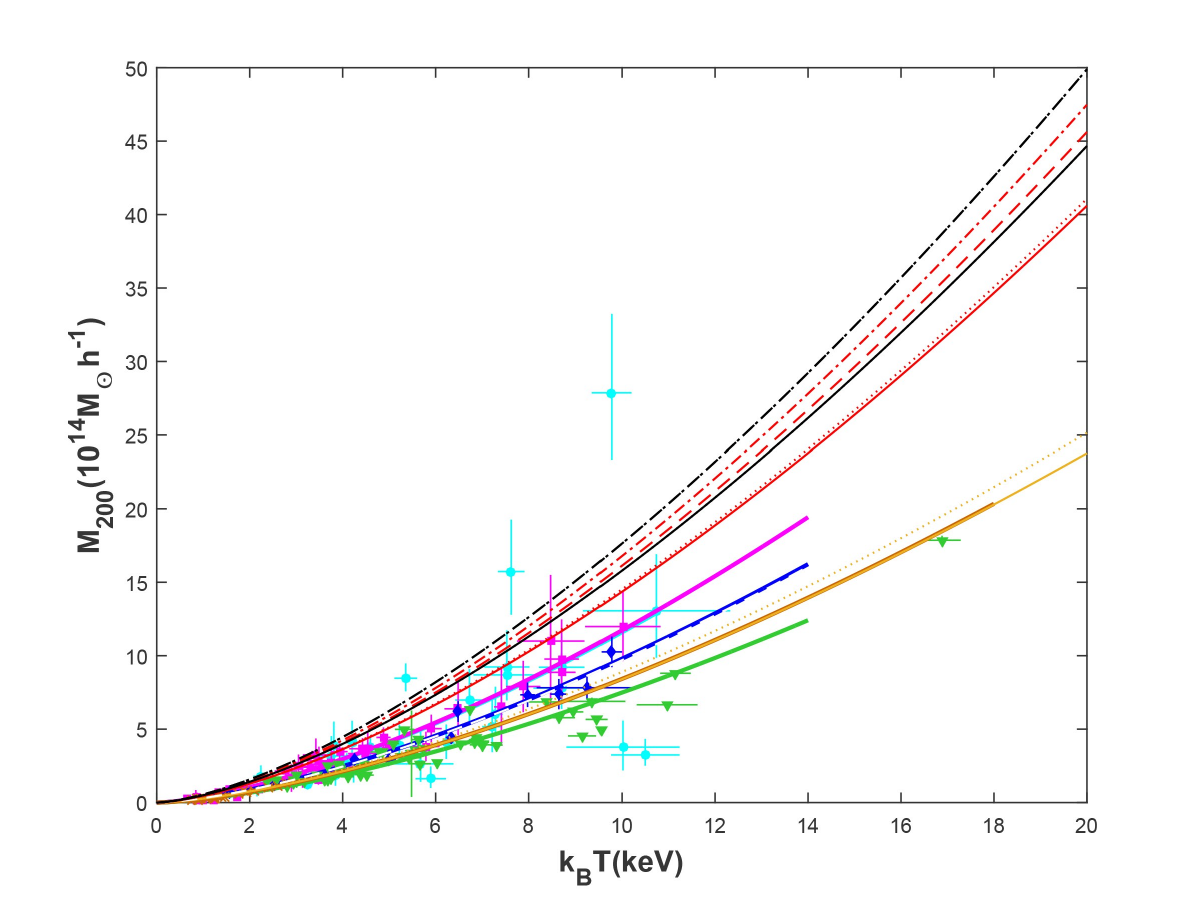}
 	\caption{The behavior of M-T relation for the predicted values of Model IV. Yellow lines represent the observations of "$SNe \, Ia + H_0$" and the other colors and types of lines are chosen completely the same as the previous M-T graphs.}
 	\label{figure_5}
 \end{figure}

For Model IV, the result of "$SNe \, Ia + H_0 + CC$" ($\xi_3=-0.27$) is impossible to indicate an actual illustration of M-T relation, while the outcomes of "$SNe \, Ia + H_0$" ($\xi_3=-0.23$), although just for NFW profile, in addition to the results of "$SNe \, Ia + H_0 + CC + BAO$" ($\xi_3=-0.038$) and "$Planck \, TT$" ($\xi_3=-1.36\times10^{-6}$) are credible. Fig. (\ref{figure_5}) represents these three predictions and compares them with observational data sets. In this graph, the prediction of "$SNe \, Ia + H_0$" is displayed by yellow lines and its "First Possibility" of NFW density profile is virtually consistent with Obs. 2015. \\

The evolution of $\lambda$ as a function of $\xi_i$ (with $i=II, III, IV$) for Models II, III, and IV are presented in Fig. (\ref{Comparison of Model II, III, and IV}). The brown, green, and magenta lines are related to Models II, III, and IV, respectively. Likewise Fig. (\ref{Comparison Model I}), black lines describe the obtained value from Obs. 1999 in which the solid lines are drawn for the "First Possibility" and dotted lines show the "Second Possibility". It demonstrates that while the $\lambda$ gradually decreases with the growth of $\xi_i$ in Models III and IV, it sharply falls for Model II. \\

\begin{figure}[h]
 	\centering
 	\includegraphics [width=0.63 \linewidth] {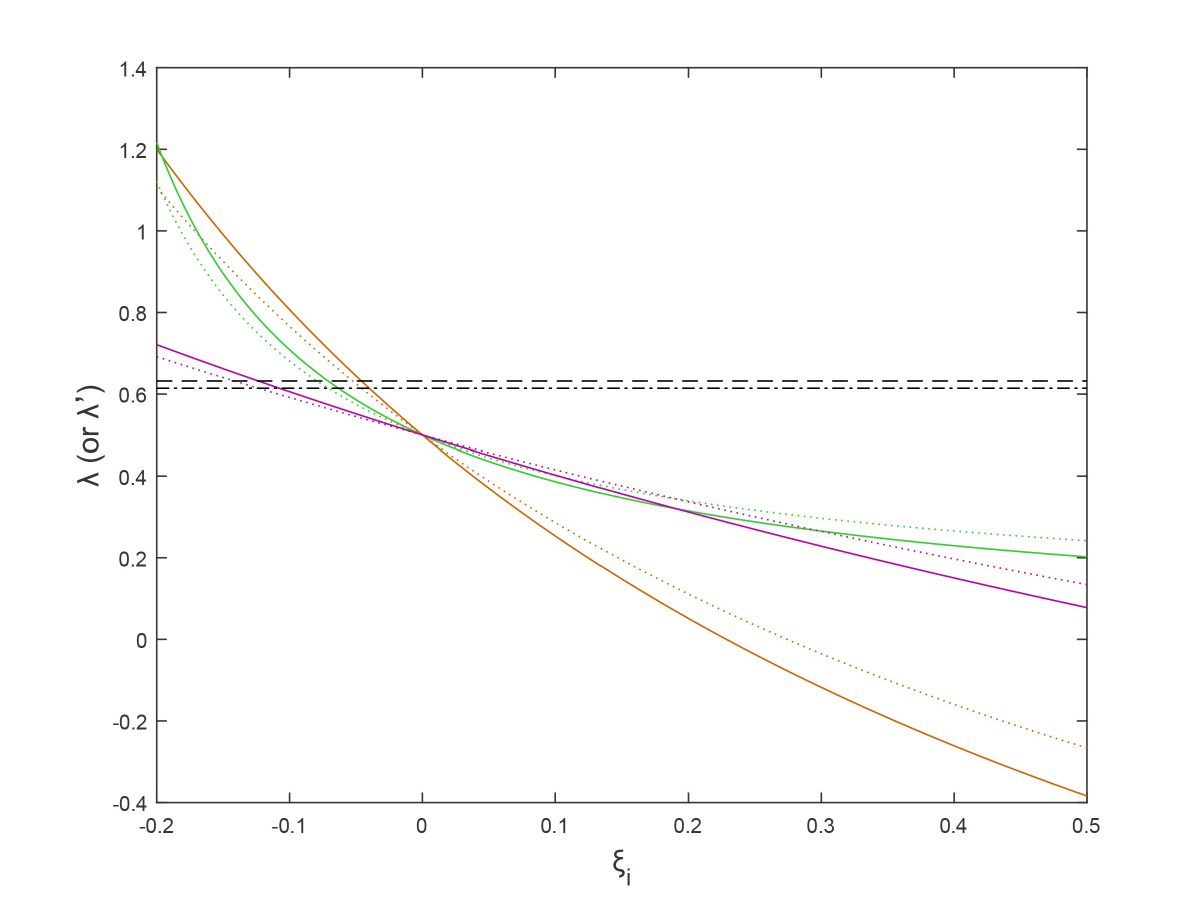}
 	\caption{The changes of $\lambda$ or $\lambda'$ as a function of interacting constant in Models II, III, and IV. The black dashed line and black dash-dot line display the "First" and the "Second" possibilities for Obs. 1999, respectively. The behavior of the three mentioned models are shown with brown (Model II), green (Model III), and magenta (Model IV) lines.}
 	\label{Comparison of Model II, III, and IV}
 \end{figure}

\begin{figure}[h]
 	\centering
 	\includegraphics [width=0.63 \linewidth] {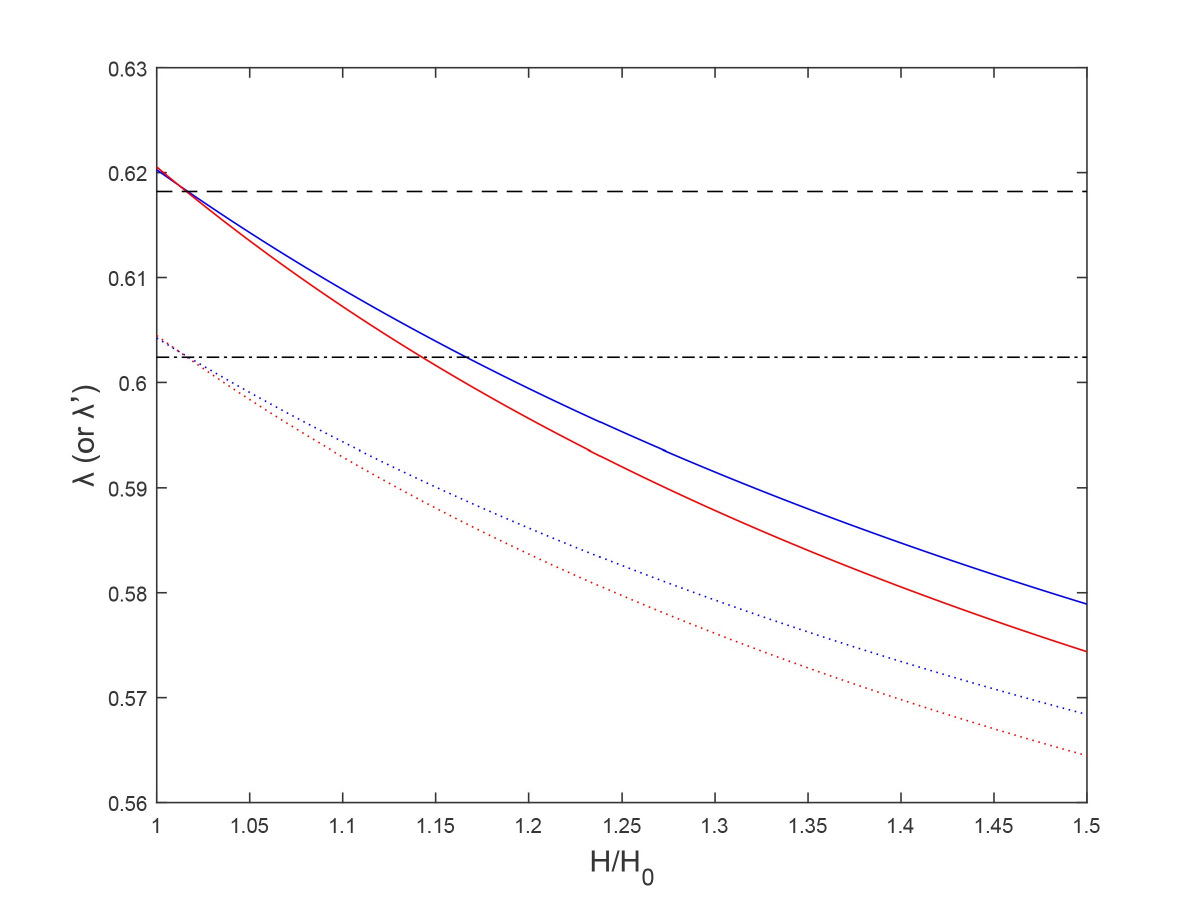}
 	\caption{The figure demonstrates how $\lambda$ and $\lambda'$ evolve with time, considering Model V. Red and blue lines are related to $\Gamma_x=0$ and $\Gamma_c=0$, respectively, and black lines denote the result of Obs. 2001 (with dashed line and solid lines representing the "First Possibility", and dash-dot line and dotted lines standing for the "Second Possibility").}
 	\label{Comparison Model V}
 \end{figure}

Model V is the most complicated one. In addition to the fact that there are two interacting parameters, there is also an important dependency on $H$ (and therefore redshift $z$), which means that $\lambda$ evolves with time. Although  \cite{vonMarttens:2018iav} does not investigate model V,  \cite{CalderaCabral:2008bx} claims that $\Gamma_x$ and $\Gamma_c$ should have opposite signs. As a second condition, it is possible to use Eq. (\ref{constraint_main}) to constrain interacting constants. Fig. (\ref{Comparison Model V}) shows the evolution of $\lambda$ with time, for the simple cases of $\Gamma_x=0$ or $\Gamma_c=0$. The horizontal axis indicates $\frac{H(z)}{H_0}$ from the present time to approximately $z=0.75$, when $\frac{H(z)}{H_0}=1.5$. The blue lines are related to the case of $\Gamma_x=0$, and the red lines describe the case of $\Gamma_c=0$. Solid lines and dotted lines denote the first and the second possibilities, respectively. As an observational example, we used the result from Obs. 2001, regarding the first (black dashed line) and the second (black dash-dot line) possibilities. According to this graph, the further the cluster is located, the more noticeable difference between the cases of $\Gamma_x=0$ and $\Gamma_c=0$ can be seen. All the lines are consistent with observational data in a low-redshift, since we fixed the value of interacting constants with regard to this observational data set itself, so it is not an interesting point. In addition, the figure clearly reveals that the constant of the virial condition was much lower than its present value in the past. It means that further clusters in interacting Model V must behave more similarly to the non-interacting model. \\

The core of our work is to determine interacting constants with respect to observational data sets for mass and temperature of galaxy clusters. As it has been mentioned before, we tried to find $\lambda_i$ (and $\lambda_i'$) in a way that the M-T relation could accurately fit the observational curves. Tables (\ref{table_1}) and (\ref{table_2}) summarize the information which has been obtained for all situations, including NFW and Burkert density profiles, the first and the second possibilities, five observational data sets and seven preferred and discussed cases of Models I to V. As far as the constants are concerned, we obtained negative values for all of them. Our results are in agreement with  \cite{vonMarttens:2018iav} in terms of obtaining negative values for these constants. It means that energy transfer occurs from DM to DE. \\

In Model V, it is common to define the dimensionless constants $\gamma_j=\frac{\Gamma_j}{H_0}$ and write $\lambda_V$ as:
\begin{equation} \label{gamma} 
\lambda_V=\frac{\frac{H(z)}{H_0}-6\gamma_c}{2\frac{H(z)}{H_0}+3\gamma_c+3\gamma_x/R} \, .
\end{equation}
Therefore, we calculated the constants $\gamma_j$ rather than $\Gamma_j$. \\

From the calculated constants, it can be concluded that more negative values are needed for a cored density profile (Burkert) than a cuspy profile (NFW) to be consistent with each observational data set. \\

Note that even fine differences among observational results may play considerable roles in the calculated constants. In fact, in our method, every input value in Eq. (\ref{Q}) contributes to measuring interacting constants. However, we strove to incorporate as many various assumptions as possible (embracing different density profiles, different possibilities, and different observational data sets) in order to compensate for the inaccuracies of parameters within $Q$.

\begin{table}[h]
\caption{The calculated constants of interacting models regarding the "First Possibility", based on making comparison with observational data sets of mass and temperature in galaxy clusters.}
\begin{tabular}{cc|c|c|c|c|c|c|c|}
\cline{3-9}
\multicolumn{1}{l}{}                         & \multicolumn{1}{l|}{} & \begin{tabular}[c]{@{}c@{}}$\alpha_c$\\ $(\alpha_x=0)$\end{tabular} & \begin{tabular}[c]{@{}c@{}}$\alpha_x$\\ $(\alpha_c=0)$\end{tabular} & $\xi_1$  & $\xi_2$  & $\xi_3$  & \begin{tabular}[c]{@{}c@{}}$\gamma_c$\\ $(\gamma_x=0)$\end{tabular} & \begin{tabular}[c]{@{}c@{}}$\gamma_x$\\ $(\gamma_c=0)$\end{tabular} \\ \hline
\multicolumn{1}{|c|}{\multirow{5}{*}{NFW}}   & Obs. 1999             & $-0.0334$                                                         & $-0.0517$                                                          & $-0.0459$ & $-0.0709$ & $-0.1235$ & -                                                               & -                                                               \\ \cline{2-9} 
\multicolumn{1}{|c|}{}                       & Obs. 2001             & $-0.0301$                                                          & $-0.0472$                                                         & $-0.0412$ & $-0.0647$ & $-0.1110$ & $-0.0306$                                                          & $-0.0481$                                                          \\ \cline{2-9} 
\multicolumn{1}{|c|}{}                       & Obs. 2006             & $-0.0547$                                                          & $-0.0764$                                                          & $-0.0750$ & $-0.1048$ & $-0.2019$ & $-0.0577$                                                          & $-0.0806$                                                          \\ \cline{2-9} 
\multicolumn{1}{|c|}{}                       & Obs. 2009             & $-0.0891$                                                          & $-0.1078$                                                          & $-0.1222$ & $-0.1478$ & $-0.3290$ & $-0.0919$                                                          & $-0.1111$                                                          \\ \cline{2-9} 
\multicolumn{1}{|c|}{}                       & Obs. 2015             & $-0.0653$                                                          & $-0.0871$                                                          & $-0.0896$ & $-0.1195$ & $-0.2412$ & $-0.0660$                                                          & $-0.0881$                                                          \\ \hline
\multicolumn{1}{|c|}{\multirow{5}{*}{Burkert}} & Obs. 1999             & $-0.0519$                                                          & $-0.0734$                                                          & $-0.0711$ & $-0.1007$ & $-0.1915$ & -                                                               & -                                                               \\ \cline{2-9} 
\multicolumn{1}{|c|}{}                       & Obs. 2001             & $-0.0488$                                                          & $-0.0701$                                                          & $-0.0670$ & $-0.0961$ & $-0.1803$ & $-0.0497$                                                          & $-0.0713$                                                          \\ \cline{2-9} 
\multicolumn{1}{|c|}{}                       & Obs. 2006             & $-0.0747$                                                          & $-0.0958$                                                          & $-0.1024$ & $-0.1314$ & $-0.2758$ & $-0.0788$                                                          & $-0.1010$                                                          \\ \cline{2-9} 
\multicolumn{1}{|c|}{}                       & Obs. 2009             & $-0.1117$                                                          & $-0.1242$                                                          & $-0.1532$ & $-0.1703$ & $-0.4125$ & $-0.1152$                                                          & $-0.1280$                                                          \\ \cline{2-9} 
\multicolumn{1}{|c|}{}                       & Obs. 2015             & $-0.0880$                                                          & $-0.1070$                                                          & $-0.1207$ & $-0.1467$ & $-0.3251$ & $-0.0890$                                                          & $-0.1081$                                                          \\ \hline
\end{tabular}
\label{table_1}
\end{table}

\begin{table}[h]
\caption{The calculated constants of interacting models regarding the "Second Possibility", based on making comparison with observational data sets of mass and temperature in galaxy clusters.}
\begin{tabular}{cc|c|c|c|c|c|c|c|}
\cline{3-9}
\multicolumn{1}{l}{}                         & \multicolumn{1}{l|}{} & \begin{tabular}[c]{@{}c@{}}$\alpha_c$\\ $(\alpha_x=0)$\end{tabular} & \begin{tabular}[c]{@{}c@{}}$\alpha_x$\\ $(\alpha_c=0)$\end{tabular} & $\xi_1$  & $\xi_2$  & $\xi_3$  & \begin{tabular}[c]{@{}c@{}}$\gamma_c$\\ $(\gamma_x=0)$\end{tabular} & \begin{tabular}[c]{@{}c@{}}$\gamma_x$\\ $(\gamma_c=0)$\end{tabular} \\ \hline
\multicolumn{1}{|c|}{\multirow{5}{*}{NFW}}   & Obs. 1999             & $-0.0292$                                                         & $-0.0461$                                                          & $-0.0400$ & $-0.0632$ & $-0.1077$ & -                                                               & -                                                               \\ \cline{2-9} 
\multicolumn{1}{|c|}{}                       & Obs. 2001             & $-0.0262$                                                          & $-0.0420$                                                         & $-0.0360$ & $-0.0576$ & $-0.0968$ & $-0.0267$                                                          & $-0.0427$                                                          \\ \cline{2-9} 
\multicolumn{1}{|c|}{}                       & Obs. 2006             & $-0.0479$                                                          & $-0.0691$                                                          & $-0.0657$ & $-0.0947$ & $-0.1769$ & $-0.0506$                                                          & $-0.0729$                                                          \\ \cline{2-9} 
\multicolumn{1}{|c|}{}                       & Obs. 2009             & $-0.0786$                                                          & $-0.0992$                                                          & $-0.1078$ & $-0.1360$ & $-0.2903$ & $-0.0811$                                                          & $-0.1023$                                                          \\ \cline{2-9} 
\multicolumn{1}{|c|}{}                       & Obs. 2015             & $-0.0574$                                                          & $-0.0792$                                                          & $-0.0787$ & $-0.1086$ & $-0.2118$ & $-0.0580$                                                          & $-0.0801$                                                          \\ \hline
\multicolumn{1}{|c|}{\multirow{5}{*}{Burkert}} & Obs. 1999             & $-0.0454$                                                          & $-0.0663$                                                          & $-0.0623$ & $-0.0909$ & $-0.1677$ & -                                                               & -                                                               \\ \cline{2-9} 
\multicolumn{1}{|c|}{}                       & Obs. 2001             & $-0.0427$                                                          & $-0.0631$                                                          & $-0.0586$ & $-0.0866$ & $-0.1578$ & $-0.0435$                                                          & $-0.0642$                                                          \\ \cline{2-9} 
\multicolumn{1}{|c|}{}                       & Obs. 2006             & $-0.0657$                                                          & $-0.0875$                                                          & $-0.0901$ & $-0.1200$ & $-0.2427$ & $-0.0693$                                                          & $-0.0923$                                                          \\ \cline{2-9} 
\multicolumn{1}{|c|}{}                       & Obs. 2009             & $-0.0990$                                                          & $-0.1154$                                                          & $-0.1358$ & $-0.1582$ & $-0.3657$ & $-0.1021$                                                          & $-0.1189$                                                          \\ \cline{2-9} 
\multicolumn{1}{|c|}{}                       & Obs. 2015             & $-0.0777$                                                          & $-0.0984$                                                          & $-0.1065$ & $-0.1349$ & $-0.2868$ & $-0.0785$                                                          & $-0.0995$                                                          \\ \hline
\end{tabular}
\label{table_2}
\end{table}

\section{Conclusion}

We investigated the mass-temperature relation of galaxy clusters for a number of interacting models of dark matter and dark energy, which are summarized in Eq. (\ref{Models}). First of all, we expanded the method provided in \cite{He:2009mz} to derive the modified virial theorem for all these models of the interacting dark sector in Section II. It immediately suggested that there might be two different possibilities for this condition, regarding two plausible behaviors of dark matter through baryonic matter. Then we used the modified virial condition to obtain M-T relation with respect to three different procedures in Section III. It revealed that the effect of interaction only emerges within the normalization factor of the M-T relation. \\

The M-T relation led to a new constraint on interacting constants, which totally depends on the concentration parameter and density profile of the clusters (Eq. (\ref{constraint_main})). This constraint is used to check the suggested constants of interacting and showed that many of those suggested values are not acceptable, due to resulting in negative masses for given temperatures. \\ 

To analyze the obtained M-T relation, we focused on five different observational data sets and compared their fitted lines with many suggested values for interacting constants. We considered two outstanding density profiles, which are NFW and Burkert, and managed to calculate interacting constants for seven cases of the five interacting models. Overall, it appears that according to these observational data sets, energy transfer should occur from DM to DE, which leads to negative values for interacting constants. It is completely consistent with the results of \cite{vonMarttens:2018iav}, which has investigated many other observational constraints to obtain numerical values for interacting constants. Although different observations result in minuscule differences in the figures, the figures are usually near zero. Furthermore, the positive constants can solely be obtained for Models I and V, if both constants have non-zero values. It also appears that for a cored density profile, more negative constants are obtained in comparison with a cuspy profile. \\ 

In the meantime the M-T relation and interacting constants were being studied, we also allocated some parts of this paper to discuss how the ratio of kinetic to the potential energy of a virialized cluster behaves as a function of interacting constants or redshift, for many of our interacting models. Fig. (\ref{Comparison Model I}) and Fig. (\ref{Comparison of Model II, III, and IV}) show that various models of interaction cause different behaviors of $\lambda$ as a function of interacting constant, although all of them lead to decreasing functions. The graphs also indicated that for Model V, the value of $\lambda$ grows with time, resulting in the fact that more distant clusters must be theoretically more consistent with non-interacting models. Two specific cases of this model ($\Gamma_x=0$ and $\Gamma_c=0$) are also more distinguishable from each other when the cluster is located in a higher redshift. \\

Finally, we emphasized that the obtained values could be extremely affected by the other parameters in the normalization factor of the M-T relation, which we have fixed with particular values for our research. However, considering a variety of possibilities might have compensated for these unwanted errors and impacts to some extend. \\

We should also mention that future observations of cluster masses and temperatures may assist to obtain more exact numerical values for interacting constants. To this purpose, cluster masses should be determined via the other methods of mass measurements, such as gravitational lensing, instead of obtaining the mass from X-ray temperature. $Euclid$ satellite and $LSST$ are two upcoming projects which would provide improved mass data through gravitational lensing observations. To have a better temperature data set, $eROSITA$ is one of the X-ray surveys that would help. In addition, future simulations with regard to verified assumptions according to observational results can suggest improved density profile and velocity dispersion for galaxy clusters and consequently, play a beneficial role in the certainty of our calculations. The impact of velocity dispersion emerges in $\tilde{\beta}_{spec}$. Clearly, any change in the assumed characteristics of the halo profile can affect the final outcomes. \\



\begin{thebibliography}{99}

\bibitem{DM-galaxy} 
  G.~Bertone, D.~Hooper and J.~Silk,
  Phys.\ Rept.\  {\bf 405}, 279 (2005)
  doi:10.1016/j.physrep.2004.08.031.

\bibitem{cluster-ga}
E.~S.~Battistelli {\it et al.},
Int.\ J.\ Mod.\ Phys.\ D {\bf 25}, no. 10, 1630023 (2016)
doi:10.1142/S0218271816300238.

\bibitem{anisotropies}
A.~Challinor,
IAU Symp.\  {\bf 288}, 42 (2013)
doi:10.1017/S1743921312016663.

\bibitem{shear}
 M.~Kilbinger,
 Rept.\ Prog.\ Phys.\  {\bf 78}, 086901 (2015)
 doi:10.1088/0034-4885/78/8/086901.

\bibitem{structure-formation}
A.~Del Popolo,
Astron.\ Rep.\  {\bf 51}, 169 (2007)
doi:10.1134/S1063772907030018.

\bibitem{large-structure}
 J.~Einasto,
 ASP Conf.\ Ser.\  {\bf 252}, 85 (2001)
 [astro-ph/0012161].

\bibitem{large-scale}
M.~Klasen, M.~Pohl and G.~Sigl,
Prog.\ Part.\ Nucl.\ Phys.\  {\bf 85}, 1 (2015)
doi:10.1016/j.ppnp.2015.07.001.

\bibitem{lamda}
 A.~G.~Riess {\it et al.} [Supernova Search Team],
 Astron.\ J.\  {\bf 116}, 1009 (1998)
 doi:10.1086/300499.

\bibitem{cosmic-coincidence}
 A.~V.~Astashenok and A.~del Popolo,
 Class.\ Quant.\ Grav.\  {\bf 29}, 085014 (2012)
 doi:10.1088/0264-9381/29/8/085014;
   H.~E.~S.~Velten, R.~F.~vom Marttens and W.~Zimdahl,
   Eur.\ Phys.\ J.\ C {\bf 74}, no. 11, 3160 (2014)
   doi:10.1140/epjc/s10052-014-3160-4;
 S.~Weinberg,
 Rev.\ Mod.\ Phys.\  {\bf 61}, 1 (1989).
 doi:10.1103/RevModPhys.61.1.

\bibitem{Farrar:2003uw}
G.~R.~Farrar and P.~J.~E.~Peebles,
Astrophys. J. \textbf{604}, 1-11 (2004)
doi:10.1086/381728.

\bibitem{CalderaCabral:2008bx}
G.~Caldera-Cabral, R.~Maartens and L.~A.~Urena-Lopez,
Phys. Rev. D \textbf{79}, 063518 (2009)
doi:10.1103/PhysRevD.79.063518.

\bibitem{interacting-observation} 
A.~A.~Costa, X.~D.~Xu, B.~Wang and E.~Abdalla,
JCAP {\bf 1701}, 028 (2017)
doi:10.1088/1475-7516/2017/01/028;
W.~Yang, S.~Pan, E.~Di Valentino, R.~C.~Nunes, S.~Vagnozzi and D.~F.~Mota,
JCAP {\bf 1809}, 019 (2018)
doi:10.1088/1475-7516/2018/09/019;
 E.~Di Valentino, A.~Melchiorri, O.~Mena and S.~Vagnozzi,
 Phys.\ Dark Univ.\  {\bf 30}, 100666 (2020)
 doi:10.1016/j.dark.2020.100666.


\bibitem{N-body-simulations}
C.~Llinares, A.~Knebe and H.~Zhao,
Mon.\ Not.\ Roy.\ Astron.\ Soc.\  {\bf 391}, 1778 (2008)
doi:10.1111/j.1365-2966.2008.13961.x ;

G.~B.~Zhao, B.~Li and K.~Koyama,
Phys.\ Rev.\ D {\bf 83}, 044007 (2011)
doi:10.1103/PhysRevD.83.044007 ;

E.~Puchwein, M.~Baldi and V.~Springel,
Mon.\ Not.\ Roy.\ Astron.\ Soc.\  {\bf 436}, 348 (2013)
doi:10.1093/mnras/stt1575 ;

C.~Llinares, D.~F.~Mota and H.~A.~Winther,
Astron.\ Astrophys.\  {\bf 562}, A78 (2014)
doi:10.1051/0004-6361/201322412 ;

M.~B.~Gronke, C.~Llinares and D.~F.~Mota,
Astron.\ Astrophys.\  {\bf 562}, A9 (2014)
doi:10.1051/0004-6361/201322403 .

\bibitem{turnaround}
S.~Bhattacharya, K.~F.~Dialektopoulos, A.~E.~Romano, C.~Skordis and T.~N.~Tomaras,
JCAP {\bf 1707}, 018 (2017)
doi:10.1088/1475-7516/2017/07/018;
R.~C.~C.~Lopes, R.~Voivodic, L.~R.~Abramo and L.~Sodré, Jr.,
JCAP {\bf 1809}, 010 (2018)
doi:10.1088/1475-7516/2018/09/010

\bibitem{splashback}
S.~Adhikari, J.~Sakstein, B.~Jain, N.~Dalal and B.~Li,
JCAP {\bf 1811}, 033 (2018)
doi:10.1088/1475-7516/2018/11/033


\bibitem{Hammami-16-74}
A.~Hammami and D.~F.~Mota,
Astron.\ Astrophys.\  {\bf 598}, A132 (2017)
doi:10.1051/0004-6361/201629003
.

\bibitem{Afshordi:2001ze} 
N.~Afshordi and R.~Cen,
Astrophys.\ J.\  {\bf 564}, 669 (2002)
doi:10.1086/324282.

\bibitem{Popolo} 
A.~D.~Popolo,
Mon.\ Not.\ Roy.\ Astron.\ Soc.\  {\bf 336}, 81 (2002)
doi:10.1046/j.1365-8711.2002.05697.x.

\bibitem{DelPopolo:2019oxn}
A.~Del Popolo, F.~Pace and D.~F.~Mota,
Phys. Rev. D \textbf{100}, no.2, 024013 (2019)
doi:10.1103/PhysRevD.100.024013.

\bibitem{CalderaCabral:2009ja}
G.~Caldera-Cabral, R.~Maartens and B.~M.~Schaefer,
JCAP \textbf{07}, 027 (2009)
doi:10.1088/1475-7516/2009/07/027.

\bibitem{vonMarttens:2018iav}
R.~von Marttens, L.~Casarini, D.~F.~Mota and W.~Zimdahl,
Phys. Dark Univ. \textbf{23}, 100248 (2019)
doi:10.1016/j.dark.2018.10.007.

\bibitem{grvirial}
R.~Javadinezhad, J.~T.~Firouzjaee and R.~Mansouri,
Phys. Rev. D \textbf{93}, no.2, 023007 (2016)
doi:10.1103/PhysRevD.93.023007

\bibitem{grvirialMG}
T.~Harko and K.~S.~Cheng,
Phys.\ Rev.\ D {\bf 76}, 044013 (2007)
doi:10.1103/PhysRevD.76.044013; 
N.~S.~Santos and J.~Santos,
JCAP {\bf 1512}, 002 (2015)
doi:10.1088/1475-7516/2015/12/002.

\bibitem{He:2010ta}
J.~H.~He, B.~Wang, E.~Abdalla and D.~Pavon,
JCAP \textbf{12}, 022 (2010)
doi:10.1088/1475-7516/2010/12/022.

\bibitem{He:2009mz}
J.~H.~He, B.~Wang and Y.~P.~Jing,
JCAP \textbf{07}, 030 (2009)
doi:10.1088/1475-7516/2009/07/030.

\bibitem{Layzer} 
Layzer, D., 1963. A Preface to Cosmogony. I. The Energy Equation and the Virial Theorem for Cosmic Distributions. The Astrophysical Journal, 138, p.174.

\bibitem{Ade:2015fva}
P.~A.~R.~Ade \textit{et al.} [Planck],
Astron. Astrophys. \textbf{594}, A24 (2016)
doi:10.1051/0004-6361/201525833
[arXiv:1502.01597 [astro-ph.CO]].

\bibitem{Batista:2017lwf}
R.~C.~Batista and V.~Marra,
JCAP \textbf{11}, 048 (2017)
doi:10.1088/1475-7516/2017/11/048
[arXiv:1709.03420 [astro-ph.CO]].

\bibitem{Chang:2017vhs}
C.~C.~Chang, W.~Lee and K.~W.~Ng,
Phys. Dark Univ. \textbf{19}, 12-20 (2018)
doi:10.1016/j.dark.2017.10.006
[arXiv:1711.00435 [astro-ph.CO]].


\bibitem{Muanwong:2001fy}
O.~Muanwong, P.~A.~Thomas, S.~T.~Kay, F.~R.~Pearce and H.~M.~P.~Couchman,
Astrophys. J. Lett. \textbf{552}, L27 (2001)
doi:10.1086/320261.

\bibitem{Bialek}
  J.~J.~Bialek, A.~E.~Evrard and J.~J.~Mohr,
  Astrophys.\ J.\  {\bf 555}, 597 (2001)
  doi:10.1086/321507.

\bibitem{Stanek} 
R.~Stanek, E.~Rasia, A.~E.~Evrard, F.~Pearce and L.~Gazzola,
Astrophys.\ J.\  {\bf 715}, 1508 (2010)
doi:10.1088/0004-637X/715/2/1508.

\bibitem{Planelles}
S.~Planelles, S.~Borgani, D.~Fabjan, M.~Killedar, G.~Murante, G.~L.~Granato, C.~Ragone-Figueroa and K.~Dolag,
Mon.\ Not.\ Roy.\ Astron.\ Soc.\  {\bf 438}, no. 1, 195 (2014)
doi:10.1093/mnras/stt2141.

\bibitem{Burkert:1995yz}
A.~Burkert,
IAU Symp. \textbf{171}, 175 (1996)
doi:10.1086/309560
[arXiv:astro-ph/9504041 [astro-ph]].

\bibitem{Maccio:2008pcd}
A.~V.~Maccio', A.~A.~Dutton and F.~C.~v.~d.~Bosch,
Mon. Not. Roy. Astron. Soc. \textbf{391}, 1940-1954 (2008)
doi:10.1111/j.1365-2966.2008.14029.x
[arXiv:0805.1926 [astro-ph]].

\bibitem{Bhattacharya}
Bhattacharya, S., Habib, S., Heitmann, K. and Vikhlinin, A., 2013. Dark matter Halo profiles of massive clusters: Theory versus observations. The Astrophysical Journal, 766(1), p.32.

\bibitem{Lokas:2000mu}
E.~L.~Lokas and G.~A.~Mamon,
Mon. Not. Roy. Astron. Soc. \textbf{321}, 155 (2001)
doi:10.1046/j.1365-8711.2001.04007.x
[arXiv:astro-ph/0002395 [astro-ph]].

\bibitem{R88a}
Ryden, B.S., 1988. Galaxy formation-The role of tidal torques and dissipational infall. The Astrophysical Journal, 329, pp.589-611.

\bibitem{Popolo:1998fz}
A.~D.~Popolo and M.~Gambera,
Astron.\ Astrophys.\  {\bf 337}, 96 (1998).

\bibitem{Lahav:1991wc}
 O.~Lahav, P.~B.~Lilje, J.~R.~Primack and M.~J.~Rees,
 Mon.\ Not.\ Roy.\ Astron.\ Soc.\  {\bf 251}, 128 (1991).

\bibitem{Colafrancesco:1994ne}
S.~Colafrancesco, V.~Antonuccio-Delogu and A.~D.~Popolo,
Astrophys. J. \textbf{455}, 32 (1995)
doi:10.1086/176552.

\bibitem{Voit:2000ie}
G.~M.~Voit,
Astrophys. J. \textbf{543}, 113 (2000)
doi:10.1086/317084.

\bibitem{Horner:1999wf}
D.~J.~Horner, R.~F.~Mushotzky and C.~A.~Scharf,
Astrophys. J. \textbf{520}, 78-86 (1999)
doi:10.1086/307437.

\bibitem{Finoguenov:2000qb}
A.~Finoguenov, T.~H.~Reiprich and H.~Boehringer,
Astron. Astrophys. \textbf{368}, 749-759 (2001)
doi:10.1051/0004-6361:20010080.

\bibitem{Vikhlinin:2005mp}
A.~Vikhlinin, A.~Kravtsov, W.~Forman, C.~Jones, M.~Markevitch, S.~S.~Murray and L.~Van Speybroeck,
Astrophys. J. \textbf{640}, 691-709 (2006)
doi:10.1086/500288.

\bibitem{Vikhlinin:2008cd}
A.~Vikhlinin, R.~A.~Burenin, H.~Ebeling, W.~R.~Forman, A.~Hornstrup, C.~Jones, A.~V.~Kravtsov, S.~S.~Murray, D.~Nagai, H.~Quintana and A.~Voevodkin,
Astrophys. J. \textbf{692}, 1033-1059 (2009)
doi:10.1088/0004-637X/692/2/1033.

\bibitem{Lovisari:2014pka}
L.~Lovisari, T.~Reiprich and G.~Schellenberger,
Astron. Astrophys. \textbf{573}, A118 (2015)
doi:10.1051/0004-6361/201423954.


\bibitem{temperature}
B.~F.~Mathiesen and A.~E.~Evrard,
Astrophys.\ J.\  {\bf 546}, 100 (2001)
doi:10.1086/318249.

\bibitem{Horner}
D.~J.~Horner, R.~F.~Mushotzky and C.~A.~Scharf,
Astrophys.\ J.\  {\bf 520}, 78 (1999)
doi:10.1086/307437.


\end{thebibliography}
\end{document}